\documentclass[aip, pop, preprint]{revtex4-1}
\usepackage[utf8]{inputenc}

\usepackage{hyperref}
\usepackage{latexsym}
\usepackage{mathrsfs}
\usepackage{graphicx}
\usepackage{subcaption} % !!! for subfigures and others !!!
\usepackage{bm} % for /boldsymbol
\usepackage[dvipsnames]{xcolor}
\usepackage[section]{placeins}
\usepackage{amsmath}
\usepackage[normalem]{ulem}

\newcommand{\aeq}{\begin{equation}}
\newcommand{\eeq}{\end{equation}}
\newcommand{\aeqn}{\begin{eqnarray}}
\newcommand{\eeqn}{\end{eqnarray}}
\newcommand{\aeqns}{\begin{eqnarray*}}
\newcommand{\eeqns}{\end{eqnarray*}}
\newcommand{\yb}[1]{\mathbf{#1}}
\newcommand{\ybs}[1]{\boldsymbol{#1}}
\newcommand{\yfrac}[2]{\displaystyle{\frac{#1}{#2}}}

\newcommand{\ydd}[2]{\yfrac{ \diff #1}{\diff #2}}
\newcommand{\ypd}[2]{\yfrac{ \partial #1 }{\partial #2}}

\newcommand{\nwc}{[\omega_{ci}]}
\newcommand{\niwc}{[\omega^{-1}_{ci}]}
\newcommand{\nvR}{[\sqrt{2} v_{th,i}/R_0]}

\newcommand{\Rs}{\yb{\dot{R}_{sp}}}
\newcommand{\Rso}{\yb{\dot{R}_{0,sp}}}
\newcommand{\Rsp}{\yb{\dot{R}_{1,sp}}}
\newcommand{\Ef}{\mathcal{E}}
\newcommand{\pzs}{\dot{p}_{z,sp}}
\newcommand{\Bss}{\yb{B}^*_{sp}}

\newcommand{\Bpss}{B^*_{\parallel, sp}}

\newcommand{\Bpsp}{B^*_{\parallel,p}}
\newcommand*\diff{\mathop{}\!\mathrm{d}}

\newcommand{\ySubFig}[1]{
	\begin{subfigure}[t] 
		{0.49\textwidth} 
		\includegraphics[width=\textwidth]{./#1}
	\end{subfigure}
}
\newcommand{\ySubFigO}[2]{
	\begin{subfigure}[t] 
		{#2\textwidth} 
		\includegraphics[width=\textwidth]{./#1}
	\end{subfigure}
}

\newcommand{\ySubFigSL}[2]{
	\begin{subfigure}[t] 
		{0.49\textwidth} 
		\includegraphics[width=\textwidth]{./#1}\caption{\label{#2}}
	\end{subfigure}
}

\newcommand{\yFigOne}[3]{
	\begin{figure}[!ht]
		\includegraphics[width=\textwidth]{./#1}
		\caption{#2 \label{#3}}
	\end{figure}
}

\newcommand{\yFigTwo}[4]{
	\begin{figure}[!ht]
		\ySubFig{#1}\ySubFig{#2}\caption{#3 \label{#4}}
	\end{figure}
}

\newcommand{\yFigThreeSmall}[5]{
	\begin{figure}[!ht]
		\ySubFigO{#1}{0.31}\ySubFigO{#2}{0.31}\ySubFigO{#3}{0.31}
		\caption{#4 \label{#5}}
	\end{figure}
}

\newcommand{\yFigFour}[6]{
	\begin{figure}[!ht]
		\ySubFig{#1}\ySubFig{#2}
		\ySubFig{#3}\ySubFig{#4}\caption{#5 \label{#6}}
	\end{figure}
}

\begin{document}

% -------------------------------------------------------
% --- TITLE ---
% -------------------------------------------------------
\title{Implementation of energy transfer technique in ORB5 to study collisionless wave-particle interactions in phase-space.}

\author{I. Novikau}
\email[]{ivan.novikau@ipp.mpg.de}
\affiliation{Max-Planck-Institut f\"ur Plasmaphysik, 85748 Garching, Germany}

\author{A. Biancalani}
\affiliation{Max-Planck-Institut f\"ur Plasmaphysik, 85748 Garching, Germany}

\author{A. Bottino}
\affiliation{Max-Planck-Institut f\"ur Plasmaphysik, 85748 Garching, Germany}

\author{A. Di Siena}
\affiliation{Max-Planck-Institut f\"ur Plasmaphysik, 85748 Garching, Germany}

\author{Ph. Lauber}
\affiliation{Max-Planck-Institut f\"ur Plasmaphysik, 85748 Garching, Germany}

\author{E. Poli}
\affiliation{Max-Planck-Institut f\"ur Plasmaphysik, 85748 Garching, Germany}

\author{E. Lanti}
\affiliation{{\'E}cole Polytechnique F\'ed\'erale de Lausanne, Swiss Plasma Center, Switzerland}

\author{L. Villard}
\affiliation{{\'E}cole Polytechnique F\'ed\'erale de Lausanne, Swiss Plasma Center, Switzerland}

\author{N. Ohana}
\affiliation{{\'E}cole Polytechnique F\'ed\'erale de Lausanne, Swiss Plasma Center, Switzerland}

\author{S. Briguglio}
\affiliation{ENEA C.R. Frascati, Via Enrico Fermi 45, CP 65-00044 Frascati, Italy}

\date{\today}

% -------------------------------------------------------
% --- ABSTRACT ---
% -------------------------------------------------------
\begin{abstract}
A new diagnostic has been developed to investigate the wave-particle interaction in the phase-space in gyrokinetic particle-in-cell codes. 
Based on the projection of energy transfer terms onto the velocity space, the technique has been implemented and tested in the global code 
ORB5 and it gives an opportunity to localise velocity domains of maximum wave-plasma energy exchange for separate species. 
Moreover, contribution of different species and resonances can be estimated as well, by integrating the energy transfer terms in corresponding velocity domains. 
This Mode-Plasma-Resonance (MPR) diagnostic has been applied to study the dynamics of the Energetic-particle-induced Geodesic Acoustic Modes (EGAMs) in an ASDEX Upgrade shot, by analysing the influence of different species on the mode time evolution.
Since the equations on which the diagnostic is based, are valid in both linear and nonlinear cases, this approach can be applied to study nonlinear plasma effects. 
As a possible future application, the technique can be used, for instance, to investigate the nonlinear EGAM frequency chirping, or the plasma heating due to the damping of the EGAMs.
\end{abstract}


\maketitle

% -------------------------------------------------------
% --- INTRODUCTION ---
% -------------------------------------------------------
\section{Introduction}
Gyrokinetic (GK) codes have recently become standard tools for the investigation of waves and instabilities in tokamak plasmas, with frequency below the ion cyclotron frequency~\cite{Brizard07}. Although they have been traditionally considered numerically heavy, in comparison to lighter hybrid models, in the last years GK codes have become capable of providing global electromagnetic predictions of the nonlinear plasma dynamics, thanks to smart schemes improving the numerical 
performance~\cite{Hatzky07,Mishchenko17}, and to the access to high-performance computers. One advantage of using GK codes is that their model includes kinetic effects such as wave-particle resonances, which are neglected in fluid descriptions.

Wave-particle interaction, such as Landau damping, can be best detected by phase space resolving diagnostics. 
In particular, investigating collisionless energy transfer signals as a function of particle velocity, necessary details can be provided to identify dominant collisionless processes governing the damping or growth of electrostatic (ES) zonal modes, such as geodesic acoustic modes 
(GAMs)~\cite{Winsor68, Diamond05, Qiu18} or energetic-particle driven GAMs, called EGAMs~\cite{Boswell06, Fu08, Horvath16, Ido18}. 
There are different kind of techniques to investigate dynamics of modes in the phase-space.
Correlation techniques~\cite{Howes17} can be used to clarify the origin of the energy-transfer process and the nature of mechanisms that lie beyond observed mode dynamics by calculating correlations of the energy transfer terms with different fields signals. 
Conjunction diagnostics based on the measurements at different positions along the same magnetic flux tube can be used to study the integrated effect of wave-particle interactions between the two space 
points~\cite{Keiling09}. 
The conjunction studies are particularly well suited to study the waves, that are propagating along the magnetic field lines, such as shear-Alfv\'en waves.

In this work we develop a Mode-Particle-Resonance (MPR) diagnostic in the code ORB5~\cite{Jolliet07, Bottino15} to investigate energy transfer signals in velocity space in global gyrokinetic (GK) simulations. The previous version of this diagnostic gave only time evolution of the energy transfer terms, averaged over the whole phase 
space~\cite{Bottino04, Tronko16}. 
We extend it, by taking projection of these terms onto the velocity space, that gives an opportunity to investigate the contribution of different resonances in different velocity domains to the mode dynamics.
This technique is applied in global GK simulations of an experimental shot on ASDEX-Upgrade machine to study EGAMs. These modes are characterised by the oscillations of mainly toroidally symmetric global radial ES field with frequency comparable to that of the GAMs.
The energetic particles (EPs) excite the mode through the inverse Landau damping, and EPs are displaced from higher to lower energy 
range~\cite{Ido15, Osakabe14}. On the other hand, the GAMs and EGAMs are mainly damped by Landau damping. In addition to ion Landau damping, GAMs have been found to be subject to the electron Landau 
damping~\cite{Zhang10, Biancalani17, Novikau17, Ehrlacher18} as well.
Here, we show that EGAMs are also subject to electron Landau damping, which can be as important as ion Landau damping in experimentally relevant conditions.
Moreover, in these simulations the MPR diagnostic provides additional details to clarify the role of different species in the EGAM-plasma interaction.

The GAMs and EGAMs can play a significant role in the regulation of the turbulence-transport processes.
The GAMs are an oscillating branch of zonal flows~\cite{Rosenbluth98, Diamond05} (ZFs).
The ZFs can reduce the radial transport in tokamak plasma acting as a sink for the turbulence energy through the inverse energy cascading or/and by shearing plasma eddies~\cite{Manz12, Schmitz12, Chen00, Medvedeva17}. 
Contrarily, the role of the GAMs~\cite{Zonca08, Conway11, Kobayashi18, Liang18} in the turbulence suppression is still unclear and even contradictory~\cite{Scott03, Silva17}. 
It could be explained by the fact that the GAMs can transfer the energy in both directions. They can either take the energy from the turbulence, being directly excited by instabilities~\cite{Zonca08} and arising from the ZFs due to the magnetic curvature, or they can return the energy back to the instabilities~\cite{Scott03}.
Due to this complex dynamics, the role of the EGAMs in the turbulence suppression is still a subject of study~\cite{Zarzoso13, Zarzoso17, Sasaki17, MSasaki17, Biancalani18}. 
At the same time, EGAMs might play a role of an intermediate agent between the fluctuating fields and thermal plasma, by spreading fluctuating field energy to the bulk plasma through the collisionless wave-particle interaction~\cite{Osakabe14}.
In such a way, the EGAMs might be a crucial component in tokamak plasma stabilisation and be significantly helpful in the plasma heating.
Thus, investigation of EGAMs characteristics, especially in the velocity space, is necessary for precise understanding of the transport phenomena in fusion reactors, where the EPs are produced as the result of the nuclear fusion reaction or by external sources such as neutral beam injection (NBI) or ion cyclotron resonance heating (ICRH).  

The remainder of the paper is structured in the following way.
In section \ref{sec:MPR} the theoretical background and the implementation of the MPR diagnostic in ORB5 is presented. 
After that, the processing of the output signals from the diagnostic is demonstrated, and an example of an ES simulation of the GAMs in a circular magnetic configuration is given (Sec. \ref{sec:postproc}). In Sec. \ref{sec:comptheory} we show that the MPR diagnostic verifies the GAM dispersion relation.
Having discussed the technique, the experimental AUG shot \#31213 is investigated in section \ref{sec:nled} using the developed diagnostic in linear simulations. Results, calculated in ORB5, are compared with simulations of the GENE code for an ES case with adiabatic electrons in Sec. \ref{sec:gene}.

% -------------------------------------------------------
% --- FORMULATION ---
% -------------------------------------------------------
\section{Formulation and implementation of the Mode-Particle-Resonance diagnostic}\label{sec:MPR}

\subsection{Theoretical background}
The MPR diagnostic is based on the projection of energy transfer terms onto the velocity space. 
It gives an opportunity to localize velocity domains of maximum mode-plasma interactions at particular time moments or averaged in a specific time interval, that is more relevant for the modes that oscillate in time.
By integrating the energy transfer terms in a chosen velocity domain and normalizing them to the mode energy, one gets a damping or growth rate of the mode. 
More precisely, considering the case of GAMs/EGAMs,  
the theoretical background of the diagnostic can be explained using the Poynting's theorem~\cite{Brambilla98} of electromagnetics:
\aeqn
&&\ypd{\epsilon}{t} = - \ybs{\nabla} \cdot \yb{S} 
	- \sum_{sp} \yb{J}_{sp} \cdot \yb{E} \label{eq:PoyntingTheorem},\\
&&\Ef(t) = \int \epsilon \diff V,
\eeqn
where $\Ef$ is the mode energy, $\yb{S}$ is the Poynting's vector, $\yb{J}_{sp}$ is the current, produced by a species $sp$, and $\yb{E}$ is the electric field. 
%\ycg{Comment: is it an oscillating in time complex mode energy $\Ef$ or mode energy, which just is damping in time, should be taken here is still under discussion.}
By integrating over the whole real space, we get $\int \ybs{\nabla} \cdot \yb{S} \diff V = 0$ using the Gauss's theorem.
Introducing the mode complex frequency:
\aeqn
&&\hat{\omega} = \omega + i \gamma\label{eq:hatw},
\eeqn
the mode energy evolves in time as 
\aeqn
\Ef(t) = \mathcal{E}_0 \exp(- 2 i \hat{\omega} t),\label{eq:Eft}
\eeqn
from where we can get an expression for the mode complex frequency:
\aeqn
-2i\hat\omega = \frac{1}{\Ef}\ydd{\Ef}{t},
\eeqn
Finally, using Eq. \ref{eq:PoyntingTheorem}, we get an expression for the mode damping rate:
\aeqn
&&\gamma = \sum_{sp} \gamma_{sp} = - \frac{1}{2} Re\left[\left 
	\langle\frac{\mathcal{P}}{\Ef}\right \rangle_{t}
	\right],\label{eq:gamma1}\\
&&\mathcal{P} = \sum_{sp} \mathcal{P}_{sp} = 
	\sum_{sp} \int \yb{J}_{sp} \cdot \yb{E} \diff V.\label{eq:je-general}
\eeqn
Since the GAMs/EGAMs oscillate in time, it is necessary to take time averaging on several GAM/EGAM periods to get a pure damping or growth rate of these modes. 
Moreover, the GAM/EGAM electric field energy is transformed periodically to the plasma pressure perturbation. To take into account this change we consider only the envelop of the ES field energy.
Speaking about $\mathcal P$, we are dealing with the work done by the ES field on the plasma. 
Since the mode energy is taken to be only positive, with a chosen convention for signs in Eq.~\ref{eq:Eft}-\ref{eq:hatw}, a negative rate $\gamma < 0$ corresponds to a positive signal $\mathcal{P}$, indicating the energy transfer from a wave to plasma particles. 
On the other hand, a positive rate $\gamma > 0$ corresponds to the growth of the wave.

\subsection{Discretization}
To describe the implementation of the diagnostic in the code ORB5, we should start from the discretization of the plasma distribution function in the code. ORB5 is a particle-in-cell (PIC) code, where the Vlasov equation is solved using a Monte Carlo algorithm, and the Maxwell equations are solved using a finite-element method. 
At the beginning of a simulation a finite collection of initial positions in phase space is sampled by a set of numerical markers~\cite{Bottino15}. 
Every marker has a particular magnetic moment $\mu_{sp} = m_{sp} v_{\perp, sp}^2 / (2 B)$, a position in real space $\yb{R}_{sp}$, a parallel canonical momentum $p_{z,sp}$ and it is moving in a background magnetic field $\yb{B} = \yb{b} B$ with 
\aeqn
&&\Bss = \yb{B} + \frac{c p_{z,sp}}{Z_{sp}e} 
	\ybs{\nabla} \times \yb{b},\label{eq:Bss}\\
&&\Bpss = \yb{b} \cdot \Bss.\label{eq:Bpss}
\eeqn 
Here, $c$ is the speed of light, 
$m_{sp}$ and $Z_{sp}e$ are the species mass and charge, where for electrons $Z_e e = - e$ and $e$ is the absolute value of the electron charge.
Taking a phase-space position $Z = (\yb{R}_{sp}, p_{z,sp}, \mu_{sp})$ of a species marker as a random variable, the code distributes the markers in the phase space according to the initial particle distribution function $f_{0,sp}$.
It means, that each marker is a realisation of the random variable $Z$. 
For simplicity, a marker will be considered as a particle that is moving along a particular orbit defined by the following equations of motion:
%\eqs{
%\Rs = \left(\frac{p_{z,sp}}{m_{sp}} 
%	- \frac{Z_{sp}e}{m_{sp}c} J_{0, sp} A_\parallel \right) 
%	\frac{\Bss}{\Bpss} +\\
%	\frac{c}{Z_{sp}e \Bpss} \yb{b} \times
%	\left[\mu_{sp} \ybs{\nabla}B  
%		+ Z_{sp}e\ybs{\nabla}(J_{0,sp} \Psi_{sp}) \right], \label{eq:R}
%}
%\aeqn
%&&\pzs = - \frac{\Bss}{\Bpss} \cdot \left[\mu_{sp} \ybs{\nabla}B 
%	+ Z_{sp}e\ybs{\nabla}(J_{0,sp}\Psi_{sp}) \right], \label{eq:pz}\\
%&&\dot{\mu}_{sp} = 0,	
%\eeqn
\begin{align}
\Rs = &\left(\frac{p_{z,sp}}{m_{sp}} 
	- \frac{Z_{sp}e}{m_{sp}c} J_{0, sp} A_\parallel \right) 
	\frac{\Bss}{\Bpss} + \notag\\
	&\frac{c}{Z_{sp}e \Bpss} \yb{b} \times
	\left[\mu_{sp} \ybs{\nabla}B  
		+ Z_{sp}e\ybs{\nabla}(J_{0,sp} \Psi_{sp}) \right], \label{eq:R}\\
\pzs = &- \frac{\Bss}{\Bpss} \cdot \left[\mu_{sp} \ybs{\nabla}B 
	+ Z_{sp}e\ybs{\nabla}(J_{0,sp}\Psi_{sp}) \right], \label{eq:pz}\\
\dot{\mu}_{sp} = &0,	
\end{align}
which are obtained by varying a GK Lagrangian with respect to the phase-space coordinates $Z = (\yb{R}_{sp}, p_{z,sp}, \mu_{sp})$~\cite{Tronko16, Tronko18}.
The orbits are perturbed by the field perturbation
\aeqn
\Psi_{sp} = \Phi - \frac{p_{z, sp}}{m_{sp}c} A_\parallel,
\eeqn
with $\Phi$ and $A_\parallel$ being electric and parallel magnetic potential perturbations respectively, where only $\Phi$ remains in ES simulations. 
In the gyro-kinetic approximation the code deals with the dynamics of the gyrocentres, whose orbits are perturbed by the potentials, averaged in a space domain, defined by the species Larmor radius, around a marker position. 
This averaging is represented by the operator $J_{0,sp}$. 
In the drift-kinetic approximation, the potential perturbation is considered at a space point, where a marker is localised, without performing the gyro-averaging. 
In ORB5 the thermal and fast ions can be treated either gyro-kinetically or drift-kinetically, while the electrons are calculated drift-kinetically.

The time evolution of the plasma distribution function $f_{sp}$ is described by the Vlasov equation:
\aeqn
\ydd{f_{sp}}{t} = \ypd{f_{sp}}{t} + \Rs \cdot \ybs{\nabla} f_{sp} + 
		\pzs \ypd{f_{sp}}{p_{z,sp}} = 0.\label{eq:vlasov}
\eeqn
Considering perturbations of the distribution function and of the particle orbits till the first order, one can linearize the Vlasov equation:
\begin{align}
&\ypd{\delta f_{sp}}{t} + \Rso\cdot\ybs{\nabla}\delta f_{sp} + 
	\dot{p}_{0,z,sp}\ypd{\delta f_{sp}}{p_{z,sp}}  = \notag\\
	- &\left(\ypd{f_{0, sp}}{t} + \Rso\cdot\ybs{\nabla}f_{0,sp} + 
			\dot{p}_{0,z,sp}\ypd{f_{0,sp}}{p_{z,sp}}\right) \notag\\ 
	- &\left(\Rsp\cdot\ybs{\nabla}f_{0,sp} 
		+ \dot{p}_{1,z,sp}\ypd{f_{0,sp}}{p_{z,sp}}\right)
\label{eq:Vlasov-lin} 
\end{align}
Assuming that $f_{0,sp}$ is an equilibrium distribution function, it should be conserved along unperturbed particle trajectories $(\Rso, \dot{p}_{0,z,sp})$:
\aeqn
\ydd{f_{0,sp}}{t}\bigg|_0 = \ypd{f_{0, sp}}{t} + 
	\Rso\cdot\ybs{\nabla}f_{0,sp} + 
			\dot{p}_{0,z,sp}\ypd{f_{0,sp}}{p_{z,sp}} = 0.
\eeqn
In other words, the first bracket on the right hand side of Eq.~\ref{eq:Vlasov-lin} is equal to zero. Finally, the time evolution of the perturbation of the species distribution function in linear simulations is described in the following way:
\aeqn
&&\ydd{\delta f_{sp}}{t}\bigg|_0 = - \ydd{f_{0,sp}}{t}\bigg|_1,
\eeqn
where $\bigg|_1$ indicates that it is necessary to take derivatives along the perturbed parts of species orbits $(\Rsp, \dot{p}_{1,z,sp})$. Thermal species have an equilibrium distribution function in a form of the Maxwellian one:
\aeqn
&&f_{0,sp}^{therm} = \frac{n_{sp}(\psi)}{(2\pi)^{3/2} u_{th,sp}^3(\psi)} 
	\exp\left[- \frac{m_{sp}}{T_{sp}(\psi)} 
		\left( \frac{1}{2}\left(\frac{p_{z,sp}}{m_{sp}}\right)^2 + 
			\frac{\mu_{sp} B}{m_{sp}} \right)  
	\right],\\
&&u_{th,sp}(\psi) = \sqrt{\frac{T_{sp}(\psi)}{m_{sp}}},
\eeqn
where $n_{sp}(\psi)$, $T_{sp}(\psi)$ are species density and temperature profiles along the radial coordinate $\psi$, which is the poloidal flux.
A symmetric two-bumps-on-tail distribution function has been used in this work for the fast species~\cite{Biancalani14, Zarzoso14}. This distribution assumes a flat temperature profile of the fast species:
\begin{align}
f_{0,sp}^{fast} = &A_{sp}(\psi)\exp\left[
		- \frac{m_{sp}}{T_{H,sp}} 
		\left( \frac{1}{2}\left(\frac{p_{z,sp}}{m_{sp}}\right)^2 + 
			\frac{\mu_{sp} B}{m_{sp}} \right) - 
		\frac{u_{H,sp}^2}{2 T_{H,sp}}
	\right]\notag\\
	&\cosh\left(
		\frac{p_{z,sp}}{m_{sp}}\frac{u_{H,sp}}{T_{H,sp}} 
	\right),\label{eq:two-bumps}\\
A(\psi) =& \frac{n_{sp}(\psi)}{(2\pi)^{3/2} T_{H,sp}^{3/2}}
\end{align}
where $u_{H,sp}$, $T_{H,sp}$ are constant input parameters, which specify a shift and width of the bumps respectively. 

The perturbation $\delta f$ is discretized in the 
$Z = (\yb{R}, p_z, \mu)$ phase space by $N_{sp}$ markers. 
Apart of its location $Z$, every marker has a particular weight $w_{p}(t)$, which should evolve consistently with the GK Vlasov equation Eq.~\ref{eq:vlasov}.
Here, we omit the index $sp$ to simplify equations and use the index $p$, indicating that a variable is related to a particular marker.
Detailed derivation of the weight time evolution can be found in 
Ref.~\cite{Jolliet07, Jolliet10, Bottino15}.
A marker weight can be associated to a phase space volume $\Omega_{p}$ and correspondent averaged perturbation distribution function 
$\langle\delta f\rangle_{\Omega_{p}}$:
\aeqn
&&\langle\delta f\rangle_{\Omega_{p}} = 
	\frac{1}{\Omega_{p}}\int_{\Omega_{p}}\delta f\diff \Omega_{p} =
	\frac{1}{\Omega_{p}}\int_{\Omega_{p}}
	w_p\delta(\yb{R}- \yb{R}_p)\delta(p_{z}-p_{p,z})\diff\Omega_p\\
&&w_p(t) = \langle\delta f\rangle_{\Omega_p} \Omega_p,\label{eq:w}\\
&&\lim_{\Omega_p \rightarrow 0}\langle\delta f\rangle_{\Omega_p} \rightarrow \delta f
\eeqn
Considering uniform spreading of the markers in real space and Maxwellian distribution in the velocity space, it can be shown~\cite{Jolliet10} that the phase space volume 
$\Omega_p$, associated to a marker $p$, is
\aeqn
\Omega_p = \frac{\Bpsp}{B} v_{\perp,p}
	(\pi \kappa_{v} u_{th}(s))^2 \int_0^1 \bar{J}(s) \diff s,
\eeqn
where $\bar{J}(s)$ is the flux-surface-averaged Jacobian, $\kappa_v$ defines maximum value of the species parallel and perpendicular velocities, normalized to a species thermal speed $u_{th}(s) = \sqrt{T/m}$, at every radial point 
$s = \sqrt{\psi/\psi_{edge}}$.

The meaning of the variable $\Omega_{p}$ can be explained proceeding directly from the Monte Carlo integration~\cite{Bottino15}. The expectation value of an arbitrary function $\zeta(\tilde Z)$ is 
\aeqn
E[\zeta(\tilde Z)] = \int \zeta(z) f(z) dz,
\eeqn
where $\tilde Z$ is a random variable, distributed according  to the function $f$. To minimize the variance of the function $\zeta$, one can chose another distribution function $g(\tilde Z)$, which does not vanish in the support of the distribution function $f$ (so-called importance sampling):
\aeqn
E[W(Z)\zeta(Z)] = \int \zeta(z) \frac{f(z)}{g(z)} g(z) dz.
\eeqn
In this case, speaking in terms of marker weights and using random variable $Z$, distributed with density $g$, the expectation value of the function $\zeta(\tilde Z)$ is calculated as
\aeqn
&&E[\zeta(\tilde Z)] = E[W(Z)\zeta(Z)] = \frac{1}{N} \sum_{i = 1}^N w(Z_i) \zeta(Z_i),\\
&&w(Z_i) = \frac{f(Z_i)}{g(Z_i)} = f(Z_i) \Omega(Z_i),
\eeqn
that is consistent with Eq.~\ref{eq:w}. 
In other words, if we have a small amount of markers in a finite phase space volume, their weights will be increased in comparison to a domain where there are higher number of markers at the same phase space volume. More details can be found in Ref.~\cite{Bottino15}.

To clarify different terms in Eq.~\ref{eq:R}, the characteristic $\Rs$ can been split on several terms:
\aeqn
&&\Rs = \yb{v}_{\parallel, sp} + \yb{v}_{\nabla B, sp} + 
	\yb{v}_{curvB, sp} + \yb{v}_{\nabla p, sp} + 
	\yb{v}_{E\times B, sp} + \yb{v}_{A_\parallel, sp},\\
&&\yb{v}_{\parallel, sp} = \frac{p_{z,sp}}{m_{sp}}\yb{b},
	\label{eq:vpar}\\
&&\yb{v}_{\nabla B, sp} = \mu_{sp} B \frac{1}{Z_{sp} e \Bpss} 
	\yb{b} \times \frac{\ybs{\nabla} B}{B},
	\label{eq:vbnB}\\
&&\yb{v}_{curvB, sp} = \left(\frac{p_{z,sp}}{m_{sp}}\right)^2 
	\frac{m_{sp}}{Z_{sp} e \Bpss}
	\yb{b} \times \frac{\ybs{\nabla} B}{B},
	\label{eq:vcB}\\
&&\yb{v}_{\nabla p, sp} = - \left(\frac{p_{z,sp}}{m_{sp}}\right)^2 
	\frac{m_{sp}}{Z_{sp}e \Bpss} \yb{b} \times 
	\left(\yb{b} \times \frac{\ybs{\nabla}\times 
	\yb{B}}{B} \right),\label{eq:vgradp}\\
&&\yb{v}_{E\times B,sp} = - \frac{\ybs{\nabla}(J_{0,sp}\Phi)
	\times \yb{b}}{\Bpss},\label{eq:vexb}\\
&&\yb{v}_{A_\parallel,sp} = \frac{p_{z,sp}}{m_{sp}}
	\frac{J_{0,sp} A_\parallel}{\Bpss} \yb{b} 
	\times (\yb{b}\times(\ybs{\nabla} \times \yb{b})) 
	- \frac{Z_{sp}e}{m_{sp}} J_{0,sp} A_\parallel \yb{b},
	\label{eq:vap}
\eeqn
where $\yb{b} \times (\ybs{\nabla}\times \yb{B}) / B = \ybs{\nabla}p/B^2$ in Eq.~\ref{eq:vgradp} indicates the dependence on the gradient of the kinetic plasma pressure $p$.
A precise form of the GK energy transfer signal, valide in both linear and nonlinear cases, can be derived from the GK Hamiltonian using the Noether theorem as it is shown in Ref.~\cite{Tronko16}:
\aeqn
\mathcal{P}_{sp} = -Z_{sp} e \int_V \diff V \int_{W_{sp}} 
	\diff W_{sp} \delta f_{sp} \Rso 
		\cdot \ybs{\nabla}(J_{0,sp} \Phi)
\label{eq:jdote-GK}
\eeqn
with $V$ and $W_{sp}$ being real and velocity spaces. 
By integrating the signal over the whole real space $V$ and in a small velocity domain $\Delta W_{sp}$, related to a particular velocity bin, we project the energy transfer signal to the velocity space of a particular species:
\begin{align}
\mathcal{P}_{sp} = 
	-\frac{Z_{sp} e}{N_{sp}} 
		\sum_{i \in V, \Delta W_{sp}} w_{i,sp} 
		(\yb{v}_{i, \parallel, sp} + \yb{v}_{i, \nabla B, sp} + \notag\\ 
		\yb{v}_{i, curvB, sp} + \yb{v}_{i, \nabla p, sp}) 
		\cdot \ybs{\nabla}(J_{0,sp} \Phi)|_i,\label{eq:jdote-disc}
\end{align}
where the sum $\sum_{i \in V, \Delta W_{sp}}$ is taken on all markers in the phase volume $V\Delta W_{sp}$. 
The gyro-averaged electric field $-\ybs{\nabla}(J_{0,sp} \Phi)|_i$ is taken at a position of a marker $i$. The sum is normalized to a total number of species markers $N_{sp}$ in the whole phase-space domain. In the current version of the diagnostic, only the electrostatic part of $\Rs$ is taken into account (Eq.~\ref{eq:vpar} - \ref{eq:vgradp}).
Since the GK model, that is used in ORB5, is based on the Hamiltonian formulation~\cite{Tronko16}, $p_{z,sp}$ is used as one of the velocity variables: 
\aeqn
p_{z,sp} = m_{sp}v_{\parallel, sp} + \frac{Z_{sp}e}{c}J_0 A_\parallel,
\label{eq:pz-vp}
\eeqn
This is a common choice in most of the modern GK PIC codes. In the MPR diagnostic a variable $u_{sp}$ is used for a parallel velocity:
\aeqn
u_{sp} = \frac{p_{z,sp}}{m_{sp}}.
\eeqn
In the ES case, the variables $u_{sp}$ and $v_{\parallel, sp}$ are identical $u_{sp} = v_{\parallel, sp}$, and in EM simulations with low $\beta$ they are close $u_{sp} \approx v_{\parallel, sp}$. 
With the rise of $\beta$, the difference between these two variables increases because of the contribution of the parallel magnetic potential $A_\parallel$.
A proper transition from the variable $p_{z,sp}$ to the variable $v_{\parallel, sp}$ (instead of $u_{sp}$) is necessary for the investigation of the dynamics of EM modes and for proper analysis of EM simulations. It is a matter of future publications.

\section{Post-processing}\label{sec:postproc}
Here, a GAM in a circular magnetic configuration is considered to show how the diagnostic is organised, and how the MPR data are treated. 
A circular deuterium plasma with flat safety factor $q = 1.5$, and flat density and temperature radial profiles is considered. 
The temperature is defined by the value of $\rho^* = 1 / 205$, where $\rho^*=\rho_s/a$, with $\rho_s=c_s/\omega_{ci}$ and  $c_s=\sqrt{T_e/m_i}$ being the sound speed and $\omega_{ci} = Z_i e B_0/ m_i$ being the ion cyclotron frequency.
The simulation has been performed with the electrostatic version of ORB5 with adiabatic electrons. Since here we are interested only in the GAM dynamics, the simulation has been done without energetic species.
Non-zonal modes, i.e. modes with toroidal numbers $n \neq 0$, have been filtered out to keep only the physics of the zonal modes. Background magnetic field at the magnetic axis is $B_0 = 2.0$ T, the minor and major radii are  $a_0 = 0.5$ m, $R_0 = 1.65$ m respectively. 
For simplicity, a radial domain $s = [0.5, 1.0]$ has been simulated.
The radial coordinate is $s = \sqrt{\psi/\psi_{edge}}$, where $\psi$ is the poloidal flux coordinate.
The real space has been discretized with $n_s = 300$ grid points along the radial direction, with $n_\chi = 64$ along poloidal direction and $n_\phi = 4$ along toroidal direction. 
A time step $dt\niwc = 10$ has been chosen, where the time is normalised to the inverse deuterium cyclotron frequency $\omega_{ci}$. The number of the ion markers is $N_i = 10^8$.
\yFigTwo
{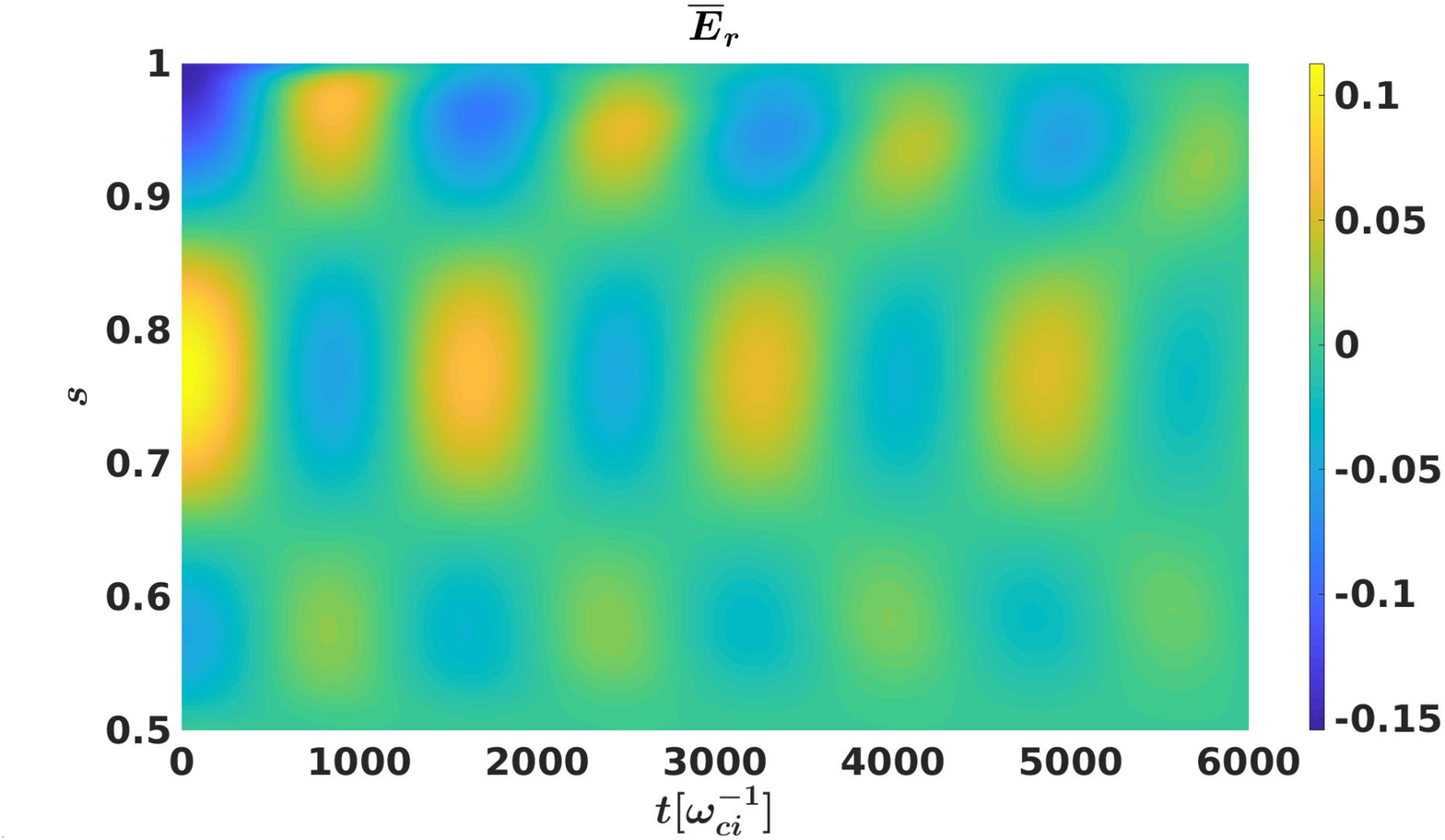}
{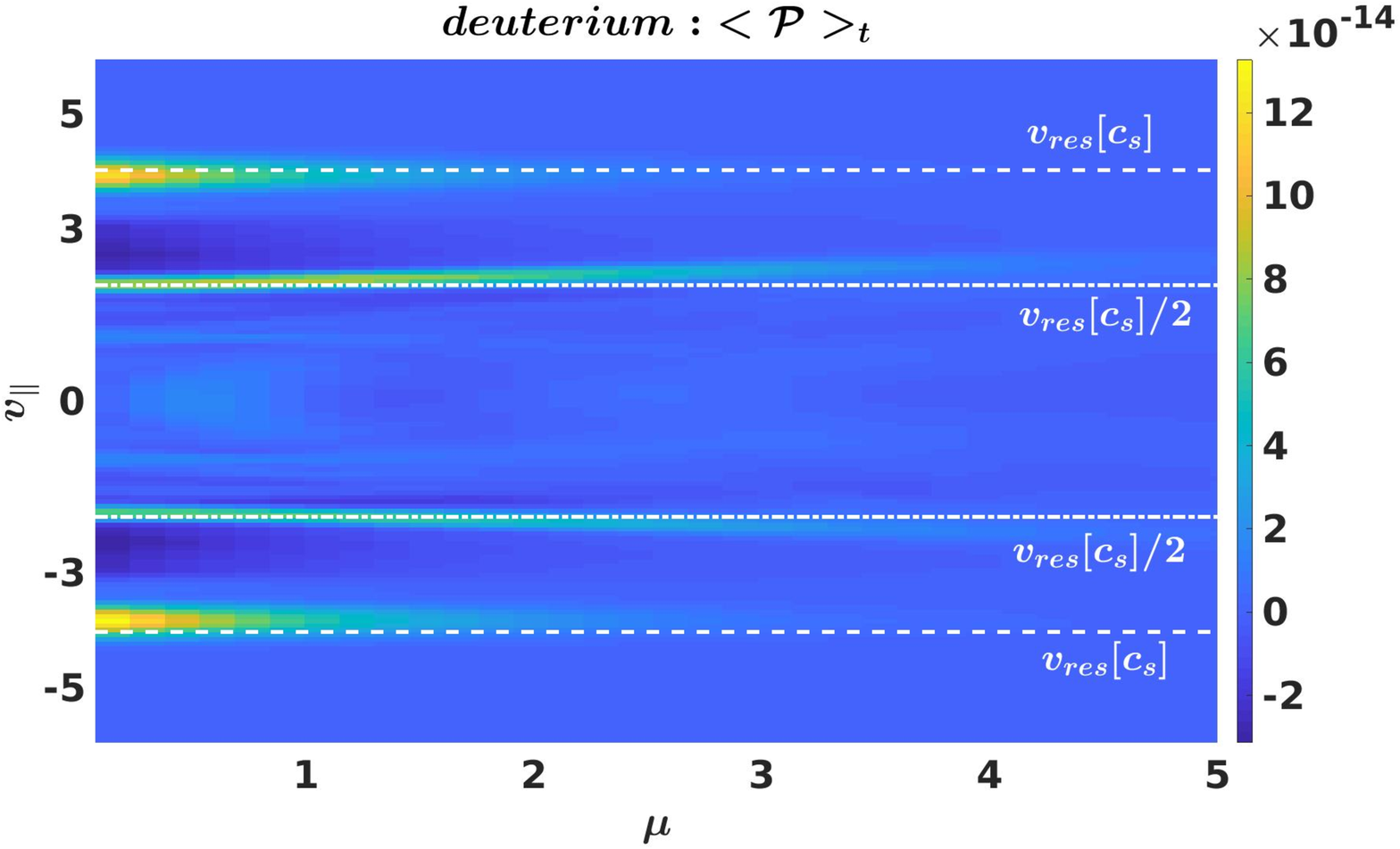}
{Time evolution of the structure of the GAM radial electric field is shown on the left plot. Velocity dependence of the energy transfer signal, averaged on several GAM periods, is shown on the right plot. White dashed and dotted lines indicate analytical estimation of the parallel velocities where the GAM-plasma resonance should be observed according to the analytical expressions Eq.~\ref{eq:vres} and Eq.~\ref{eq:vresm}. Velocity space here is normalized to the sound velocity $c_s = \sqrt{T_e(s = 0.7)/m_i}$.}
{fig:gam-er-st}
To simulate the GAM dynamics, so-called Rosenbluth-Hinton 
test~\cite{Rosenbluth98} has been sent by introducing an axisymmetric density perturbation designed to produce an initial electric potential field.

First of all, the MPR diagnostic provides the energy transfer signal $\mathcal{P}(v_\parallel, \mu, t)$ (Eq.~\ref{eq:jdote-disc}) as a function of the velocity variables $(v_\parallel, \mu)$ and time. By averaging this signal on several GAM periods, resonances of the mode-particle interaction can be localised in the velocity space. Their location can be compared with the analytically given parallel resonance velocity:
\aeqn
v_{\parallel,res} = q R_0 \omega_{GAM},\label{eq:vres}
\eeqn
where $\omega_{GAM}$ is the GAM frequency, that can be found directly from the radial zonal electric field $\overline{E}_r$. 
Since the perturbation of the plasma distribution function related to the GAM dynamics can have higher poloidal modes $m \geq 1$, the GAM-particle interaction can be observed at smaller parallel velocities as well 
\aeqn
v^{(m)}_{\parallel,res} = \frac{q R_0 \omega_{GAM}}{m}.\label{eq:vresm}
\eeqn
By integrating in corresponding velocity domains, one can estimate contribution of these resonances to the mode dynamics. 
In this particular case, it can be seen from Fig.~\ref{fig:gam-er-st}, that the energy transfer occurs mainly at the first resonance $v_{\parallel,res}$. By integrating the signal in the whole velocity domain, one gets the time evolution of $\mathcal{P}$. 
By normalizing it to the mode energy, the GAM damping rate can be estimated using Eq.~\ref{eq:gamma1}. The mode energy has been taken as an envelop (grey dotted line at the left plot in Fig.~\ref{fig:gam-mpr}) of the field energy (green line). The reason is that the GAM energy periodically oscillates between field and plasma components. 
Because of that, the total GAM energy can be estimated as an interpolation of the maxima of the ES field energy.

Eq.~\ref{eq:gamma1} involves an integration in time. Varying and choosing different time intervals, one can estimate an errorbar of the GAM damping rate by building a distribution (or histogram) of the damping rate values. 
Every chosen time interval has to contain a whole number of GAM periods. 
The result histogram can be fitted with the normal distribution function, that gives a mean value of the damping rate and a $95\%$ confidence interval as $3\sigma$, where $\sigma$ is the standard deviation, found from the distribution function. One can use a different distribution function to take into account a non-zero skewness of a histogram. For example, the  Generalized-Extreme-Value (Ref. ~\cite{Kotz00}) distribution function has been used as well. Both distribution functions give quite close results for the mean value and the standard deviation of the GAM damping rate. A result value of the GAM damping rate, found from the MPR diagnostic is the following:
\aeqn
&&\gamma\nwc = -1.1\cdot 10^{-4} \pm 4.0\cdot 10^{-5},
\label{eq:gam-mpr-gamma}
\eeqn
and the distribution of the GAM damping rate values is shown in Fig.~\ref{fig:gam-mpr}. 
%\ycg{Comment: by subtracting a zero-frequency zonal component of the radial electric field from the energy transfer signal, we can try to get a smaller errorbar for the GAM damping rate.}

\yFigTwo
{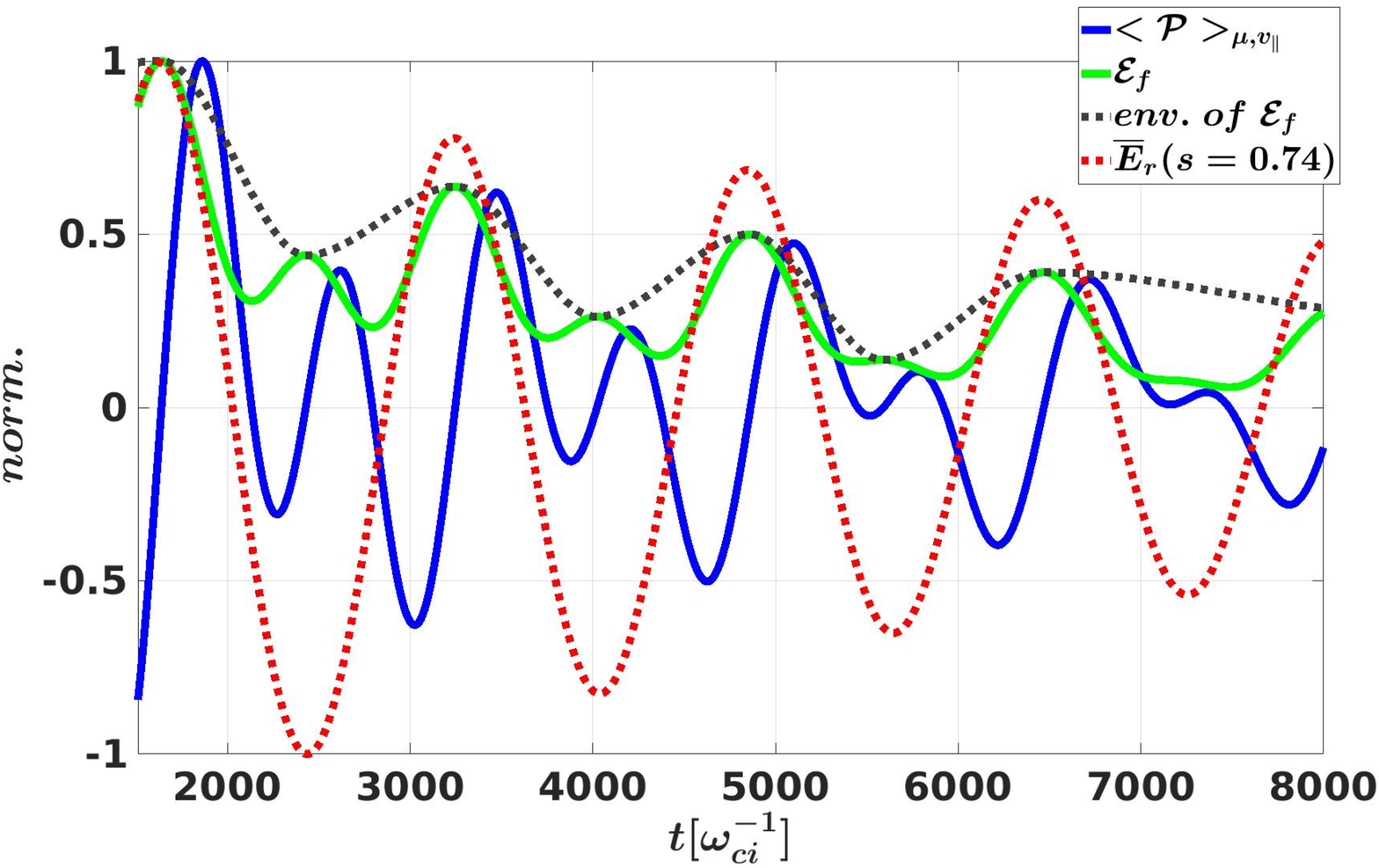}
{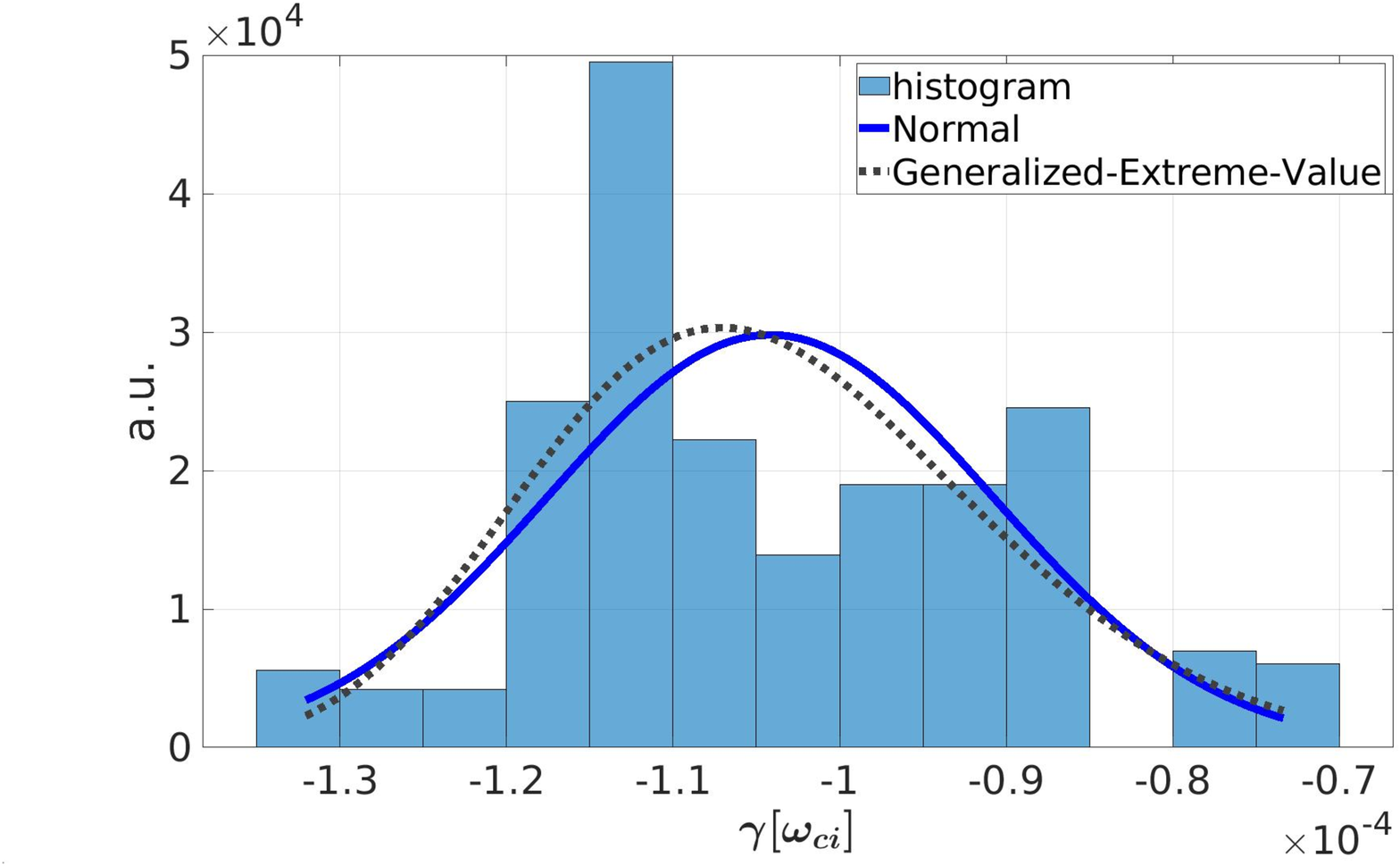}
{Time evolution of the energy transfer (blue line) and field energy signals (green line) is plotted on the left plot. Estimation of the mode energy (grey line) is taken as an envelop of the field energy.  For comparison, the zonal radial electric field at $s = 0.74$ is shown as well (red line). Distribution of the GAM damping rate, given by the MPR diagnostic, is depicted on the right plot.}
{fig:gam-mpr}

The result from the MPR diagnostic can be compared with the direct calculation of the GAM damping rate, by fitting the zonal radial electric field $\overline{E}_r$ at a particular radial point. 
Here, the point $s = 0.74$ has been taken, since it is very close to a crest in $\overline{E}_r$, that can be seen from the left plot of Fig.~\ref{fig:gam-er-st}.
First of all, a zero-frequency component of $\overline{E}_r$ is filtered out for more precise calculation of the GAM characteristics. 
After that, a GAM frequency is estimated, for example, by the Fast Fourier Transform. 
On the other hand, the damping rate is estimated by the linear least-square root method from the peaks in the time evolution of $\overline{E}_r$. This preliminary processing gives the first assumption of the GAM frequency and damping rate, that are used as initial guesses in the non-linear fitting procedure. 
A function 
\aeqn
\sim\cos(\omega t)\exp(\gamma t)\label{eq:test-function}
\eeqn
 is used as a test one, which is fitted to the time evolution of the $\overline{E}_r(s = 0.74)$.  
This method has been used previously in Ref.~\cite{Novikau17} to study the influence of the drift-kinetic electrons on the GAM dynamics in linear global GK simulations. 
But here, as in the MPR diagnostic, an opportunity to estimate errorbars of the frequency and especially of the damping (or growth) rate by varying time intervals has been added as well.

\yFigTwo
{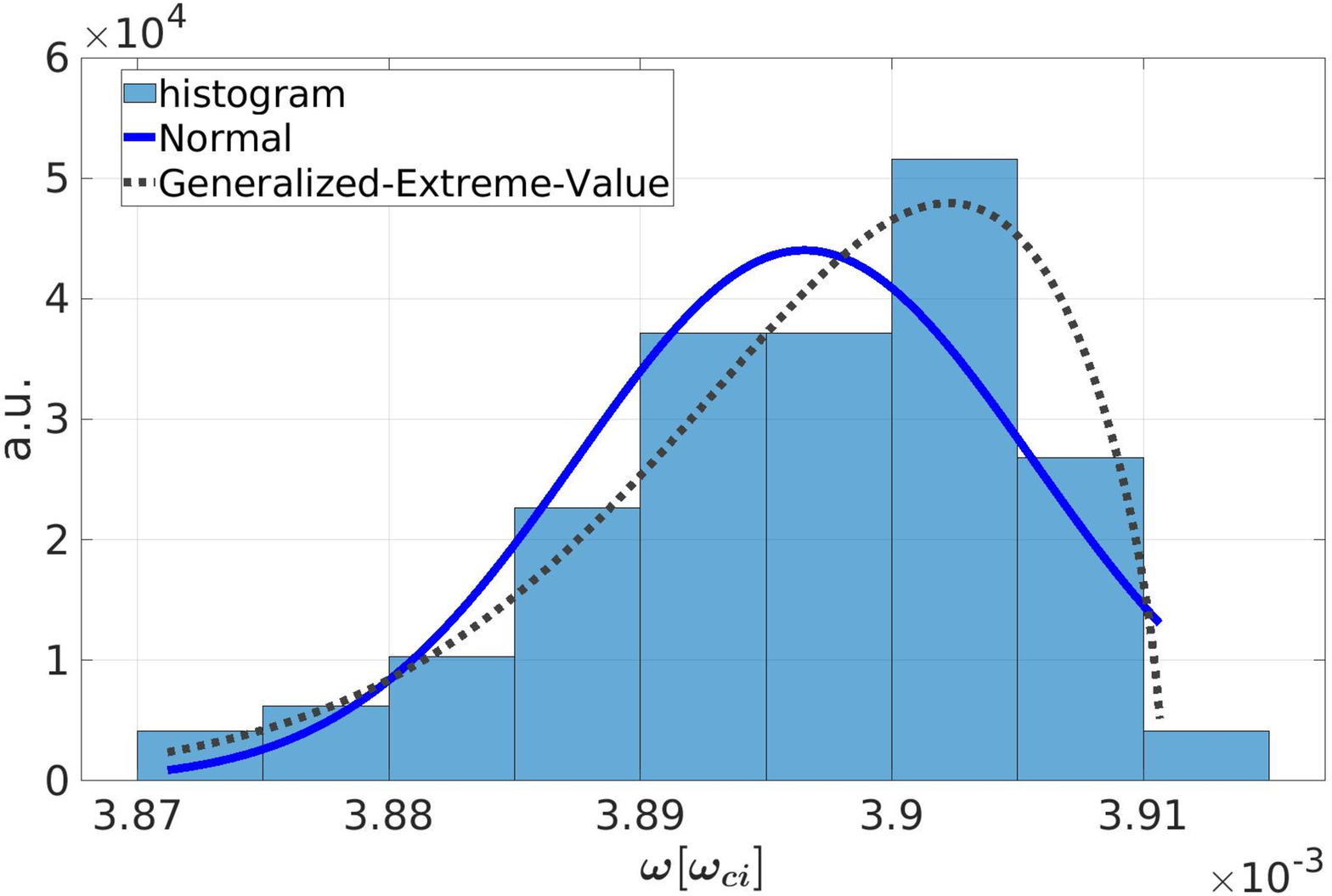}
{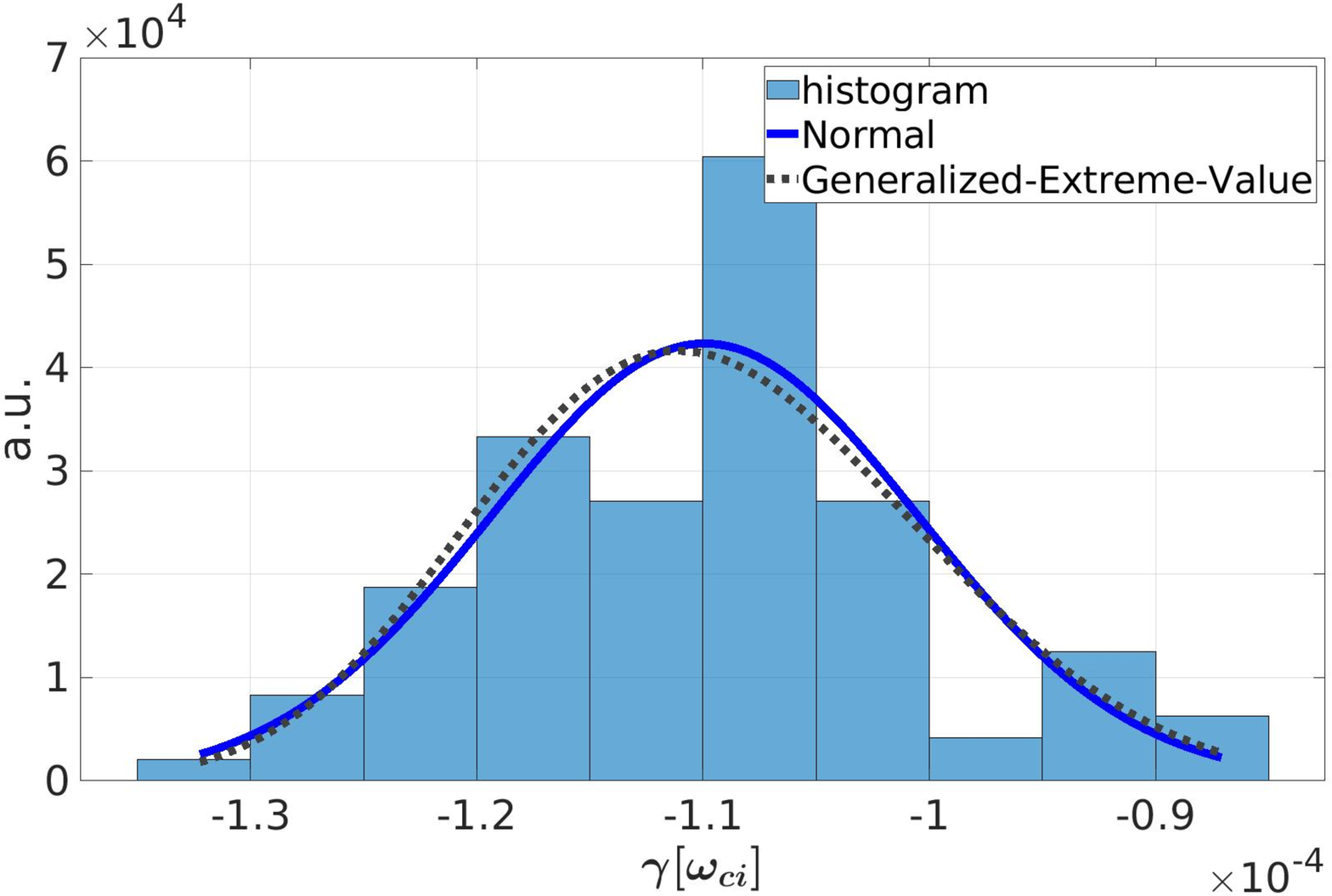}
{Distribution of the GAM frequency (left plot) and damping rate (right plot), found using the non-linear fitting of $\overline{E}_r(s1)$ to the test function Eq.~\ref{eq:test-function}.}
{fig:gam-er-wg-hists}

Finally, the GAM frequency and damping rate, found using the non-linear fitting of $\overline{E}_r(s = 0.74)$ to the test function 
Eq.~\ref{eq:test-function}, are the following:
\aeqn
&&\omega\nwc = 3.90\cdot 10^{-3} \pm 2.7\cdot 10^{-5},\label{eq:gam-freq}\\
&&\gamma\nwc = -1.1\cdot 10^{-4} \pm 2.8\cdot 10^{-5}.\label{eq:gam-gamma},
\eeqn
and have been calculated from the corresponding distribution functions, shown in Fig.~\ref{fig:gam-er-wg-hists}.
As it can be seen here, the calculation of the GAM frequency is quite precise with an errorbar being around $1\%$, while the errorbar of the damping rate prediction is around $20\%$. Comparing both methods (Eq.~\ref{eq:gam-mpr-gamma} and \ref{eq:gam-gamma}), one can see that the MPR diagnostic is not as precise as the non-linear fitting, at least, in case of the calculation of the GAM damping rate. On the other hand, it provides additional information such as a position of the GAM-plasma resonances in the velocity space (Fig.~\ref{fig:gam-er-st}).

\section{Analytical verification}\label{sec:comptheory}
Here, we are going to show the consistency of the MPR diagnostic 
by comparing the measurements on a GAM investigated with ORB5, with the analytical dispersion relation derived in the GK framework, by neglecting the effects of the finite Larmor radius and finite orbit width, and considering adiabatic electrons.
The corresponding GAM dispersion relation (Ref.~\cite{Zonca96, Zonca08}) is
\aeqn
&&z + q^2 \left(F(z) - \frac{N^2(z)}{D(z)}\right) = 0,
	\label{eq:ZoncaDisp}\\
&&N(z) = z + \left(\frac{1}{2} + z^2\right) \mathcal{Z}(z),\\
&&D(z) = \frac{1}{z}\left(1 + \frac{1}{\tau_e}\right) + 
	\mathcal{Z}(z),\\
&&F(z) = z(z^2 + \frac{3}{2}) + (z^4 + z^2 + 
	\frac{1}{2}) \mathcal{Z}(z),\\
&&\mathcal{Z}(z) = 
	\frac{1}{\sqrt{\pi}} \int^{+\infty}_{-\infty} 
	\frac{\exp(-y^2)}{y -z}\diff y,\\
&&z = \frac{\hat\omega}{\omega_t},\ 
	\omega_t = v_{th}/(q R_0),\ v_{th} = \sqrt{2T/m},
\eeqn
We omit species indices, since all relevant plasma variables are related to the deuterium. A GAM is described by the evolution of the zonal electric field:
\aeqn
&&\yb{\overline E} = (\overline E_r, 0, 0),\\
&&\overline E_r = \overline E_{r,1} \cos(k r) \exp(-i\hat\omega t),\label{eq:erbar}
\eeqn
with a radial wavenumber $k$ and a complex frequency 
$\hat\omega = \omega^{D} + i \gamma^{D}$, that verifies the dispersion relation Eq.~\ref{eq:ZoncaDisp}. 
The corresponding perturbation of the deuterium distribution function is:
\aeqn
\delta f = \frac{e}{T} 
	\frac{i \hat\omega F_0}{\hat\omega^2 - \omega_{tr}^2}
	\left(\frac{2c T}{e B_0 R_0}\frac{N(z)}{D(z)} - v_d\right)
	\overline E_r \sin{\theta_p},\label{eq:df}
\eeqn
where $c$ is the light speed, $\omega_{tr} = v_\parallel/(q R_0)$ is the passing frequency, $\theta_p$ is the poloidal angle in a simplified circular geometry, $F_0$ being the deuterium equilibrium distribution function:
\aeqn
F_0 = \left(\frac{m}{2\pi T}\right)^{3/2} 
	\exp\left(- \frac{m (v^2_\parallel + v^2_\perp)}{2 T} \right),
\eeqn
and $v_d$ being the amplitude of the radial drift composed by the curvature drift and grad-B drift:
\aeqn
v_d = \frac{m c}{e B_0 R_0}
	\left(\frac{v_\perp^2}{2} + v_\parallel^2 \right)
\eeqn
To derive an expression for the energy transfer term, we need the equation of motion  Eq. \ref{eq:R}, which in a linear ES system can be rewritten as:
\aeqn
\yb{\dot R}_0 = v_\parallel \frac{\Bss}{\Bpss} + \frac{c\mu}{e \Bpss}\yb{b} \times \ybs{\nabla} B,
\eeqn
where $\Bss$ and $\Bpss$ are defined in Eq.~\ref{eq:Bss} and \ref{eq:Bpss} respectively.
Considering a low-pressure plasma ($\yb{J}_0\times\yb{B} \ll 1$ with a plasma current $\yb{J}_0$) in a circular plasma cross-section with a curvature $\ybs{\kappa}$, one gets the following simplifications:
\aeqn
&&\ybs{\nabla}\times\yb{b} = \frac{\ybs{\nabla}\times\yb{B}}{B} + 
	\frac{\yb{B}\times\ybs{\nabla}B}{B^2} \approx 
	\frac{4\pi}{c B_0} \yb{J}_0 + \yb{b}\times\ybs{\kappa},\\
&&\yb{J}_0 \cdot \yb{\overline E} \approx 0,\\
&&\yb{b}\times\ybs{\kappa}\cdot\yb{\overline E} \approx 
	- \frac{\overline{E}_r \sin{\theta_p}}{R_0}. 
\eeqn
Applying the introduced approximations, we get that
\aeqn
\yb{\dot R}_0 \cdot \yb{E} \approx 
	\yb{\dot R}_0 \cdot \yb{\overline E} = 
		- v_d \overline E_r \sin\theta_p,
\eeqn
using which together with Eq.~\ref{eq:df} and by putting everything to Eq.~\ref{eq:jdote-GK}, an expression for the energy transfer signal can be derived:
\aeqn
&&\mathcal{P} = - \frac{e^2}{T} \int (I_1 - I_2) \overline E_r^2 
	\sin^2\theta_p \diff V,\\
&&I_1 = i\hat\omega \frac{2 c T}{e B R} \frac{N(z)}{D(z)}
	\int\frac{F_0 v_d}{\hat\omega^2 - \omega_{tr}^2}\diff W,
	\label{eq:I1}\\
&&I_2 = i\hat\omega 
	\int\frac{F_0 v_d^2}{\hat\omega^2 - \omega_{tr}^2}\diff W.
	\label{eq:I2}	
\eeqn
Evaluating the velocity integrals $I_1$ and $I_2$, one gets the expression
\aeqn
I_1 - I_2 = - i \frac{v_{th}^3}{\omega_c^2 R_0} q 
	\left(\frac{N^2(z)}{D(z)} - F(z)\right),
\eeqn
which can be significantly simplified using the GAM dispersion relation (Eq.~\ref{eq:ZoncaDisp}) to get rid of the functions $N(z), D(z)$, and $F(z)$:
\aeqn
I_1 - I_2 = -i\frac{v_{th}^2}{\omega_c^2}\hat\omega
\eeqn
As a result, we have the following complex expression for the plasma-field energy exchange signal:
\aeqn
\mathcal{P} = \int\diff V \diff W e \delta f \yb{\dot R}_0
	\cdot\yb{E}
	\approx 
	i\hat\omega\frac{2 m c^2}{B_0^2}
	\int\overline E_r^2 \sin^2\theta_p\diff V.\label{eq:je}
\eeqn
In the GK model of ORB5, the ES field energy is given (Ref.~\cite{Bottino15}) by:
\aeqn
\mathcal{E}_{mode} = \int\diff V\diff W e\delta f J_0\Phi + 
	\int\diff V\diff W F_0 
		\left(-\frac{mc^2}{2B^2}|\nabla_\perp\Phi|^2 \right)
\eeqn
with $\Phi$ being an ES potential perturbation.
Since we consider the drift-kinetic approximation here, the gyro-averaging operator $J_0$ is equal to 1. Using the GK Poisson equation, the expression can be reduced to
\aeqn
\mathcal{E}_{mode} = \frac{1}{2}\int\diff V\diff W e \delta f \Phi
\eeqn
Finally, taking into account only the perturbation of the zonal radial electric field $|\nabla_\perp \Phi|^2 \approx \overline E_r^2$ and since $\int\diff W F_0 = 1$, we get
\aeqn
\mathcal{E}_{mode} \approx \frac{mc^2}{2B_0^2}
	\int\diff V \overline E_r^2.\label{eq:Emode}
\eeqn
Since $\int\diff V \overline E_r^2 \sin\theta_p / \int\diff V \overline E_r^2 = 1/2$ and applying Eq.~\ref{eq:je}, \ref{eq:Emode} to Eq.~\ref{eq:gamma1}, the consistency between the GAM dispersion relation and the MPR method can be proved:
\aeqn
\gamma^{MPR} = - \frac{1}{2} Re\left[
	\frac{\mathcal{P}}{\mathcal{E}_{mode}}\right] = 
	- Re\left[i\hat\omega\right] = \gamma^{D}.\label{eq:consistency}
\eeqn
The expression Eq.~\ref{eq:consistency} means that using the field and plasma perturbations (Eq.~\ref{eq:erbar} and Eq.~\ref{eq:df}), which verify the GAM dispersion relation Eq.~\ref{eq:ZoncaDisp}, in the MPR diagnostic (Eq.~\ref{eq:gamma1}), one gets a GAM damping rate that verifies the starting GAM dispersion relation.

To check the time behaviour of the analytical energy transfer signal, one should take the real part of Eq.~\ref{eq:je}:
\aeqn
Re[\mathcal P] \sim 
	(\omega \sin(2\omega t) - \gamma \cos(2\omega t))\exp(2\gamma t).
	\label{eq:je_real1}
\eeqn
Taking into account, that all signals in the code ORB5 are real, we should also consider the time behaviour of the energy transfer, got from only real parts of the plasma perturbation and zonal radial electric field:
\aeqn
\mathcal{P}_{real} \sim 
	Re[\delta f] Re[\yb{\dot R}_0 \cdot \yb{\overline E}_r]
	\sim 
	(\omega \sin(2\omega t) - \gamma - 
		\gamma\cos(2\omega t))\exp(2\gamma t).
	\label{eq:je_real2}
\eeqn
Since the GAM damping rate is a small value in comparison with the GAM frequency, the Eq.~\ref{eq:je_real1} and Eq.~\ref{eq:je_real2} give the same time evolution. 
On the other hand, we should take into account the zero-frequency component of the zonal radial electric field:
\aeqn
\overline E_r = \overline E_{r,0} + 
	\overline E_{r,1} \cos(k r) \exp(-i\hat\omega t).
\eeqn
In this case, the analytical energy transfer signal takes the following form:
\begin{align}
\mathcal{P}^{ZF}_{real} \sim 
	\frac{E_{r, 0}}{E_{r,1}} (\omega \sin(\omega t) - 
		\gamma \cos(\omega t)) \exp(\gamma t)) +\notag\\
	(\omega \sin(2\omega t) - \gamma - 
		\gamma\cos(2\omega t))\exp(2\gamma t).
	\label{eq:je_real2_zf}
\end{align}

If we take the GAM frequency (Eq.~\ref{eq:gam-freq}) and damping rate (Eq.~\ref{eq:gam-mpr-gamma} or \ref{eq:gam-gamma}) from the simulation, described in Sec. \ref{sec:postproc}, we can see that Eq.~\ref{eq:je_real1} and especially \ref{eq:je_real2_zf} give a quite similar behaviour in time (Fig.~\ref{fig:theory-je}) with the same frequency of $\mathcal{P}$ as the one from the numerical simulation. And the frequency of $\mathcal P$ signal is double of that of $\overline{E}_r$.
The amplitude modulation, that is observed in the numerical energy transfer signal, can be explained by the constant component of the zonal radial electric field, that is emphasized in Eq.~\ref{eq:je_real2_zf}, and as one can see from the comparison of the blue and green lines in Fig.~\ref{fig:theory-je}.

\yFigOne
{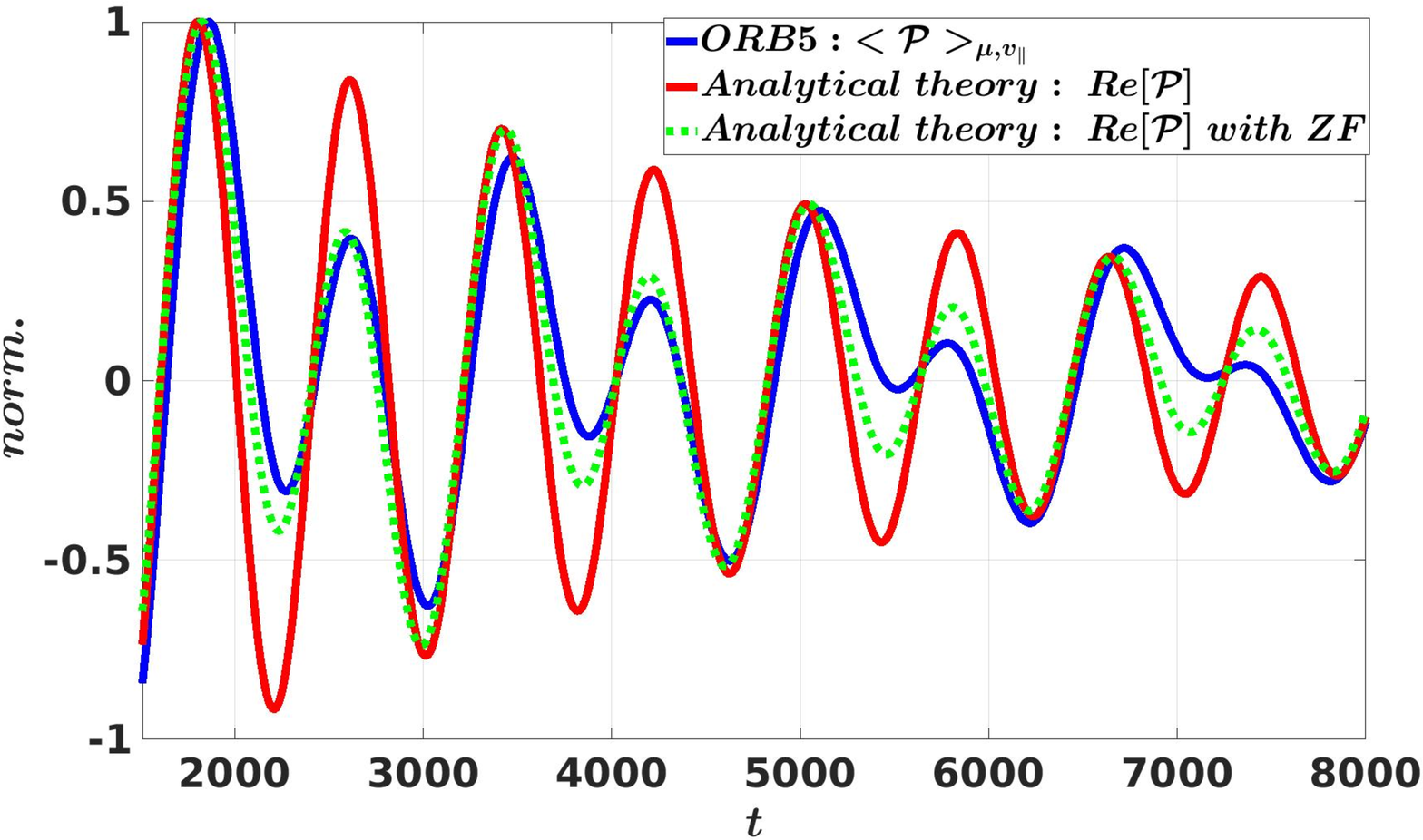}
{Comparison of the time evolution of the energy transfer signals, obtained from the numerical simulation from Sec. \ref{sec:postproc} (blue line), and analytically from Eq.~\ref{eq:je_real1} (red line) and Eq.~\ref{eq:je_real2_zf} (green line). Here, the ratio $E_{r,0}/E_{r,1} \approx 0.25$, which is used in Eq.~\ref{eq:je_real2_zf}, is estimated from the $\overline E_r(s = 0.74)$, given by the numerical simulation in ORB5.}
{fig:theory-je}

\section{Application to EGAMs in AUG shot \#31213}
\label{sec:nled}

\subsection{Equilibrium and definition of the numerical simulation}
The AUG shot \#31213 at time 0.84 s has been selected within the Non-Linear Energetic-particle Dynamics (NLED) Eurofusion enabling research project~\cite{LauberSitee, Horvath16}.  
It has been chosen to study the effect of the energetic particles (EPs) on the dynamics of EGAMs. 
That is why, in these simulations we have three species:
gyro-kinetic thermal deuterium, gyro-kinetic energetic (fast) deuterium, and thermal electrons, either adiabatic (AE) or drift-kinetic (KE). 
The linear dynamics of EGAMs in this NLED-AUG case has been recently investigated with the gyrokinetic codes GENE and ORB5 by considering adiabatic electrons~\cite{DiSiena_NF2018}. Here, we extend the previous study by investigating the effect of kinetic electrons and describing the contribution of the resonances of all species in phase space.
The simulation with the AE is performed in the electrostatic limit, while the simulation with the KE has been done including dynamics of the magnetic potential perturbation as well. 
In this latter case the pullback method~\cite{Mishchenko17} has been used for the mitigation of the cancellation problem in EM 
simulations~\cite{Chen01, Hatzky07}.
Corresponding profiles of the safety factor, species density and temperature are shown in Fig.~\ref{fig:nled-equil}. 
The magnetic field is reconstructed with experimental data, including all geometrical effects (Fig.~\ref{fig:nled-equil}).
The magnetic field at the magnetic axis is $B_0 = 2.2$ T.
The major radius at the axis is $R_0 = 1.67$ m. The geometrical major and minor radii are $R_0 = 1.62$ m, $a = 0.482$ m respectively.
\yFigFour
{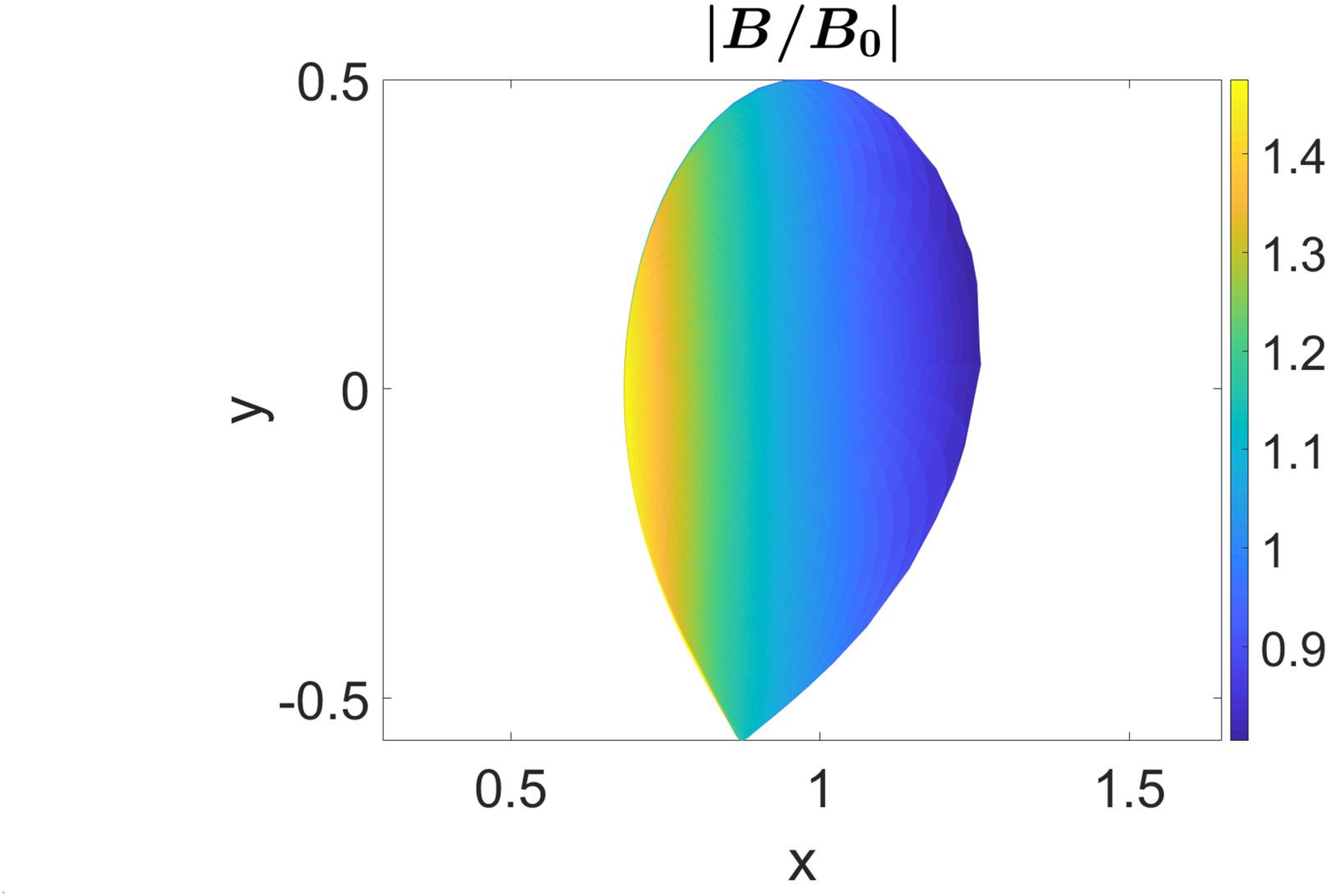}
{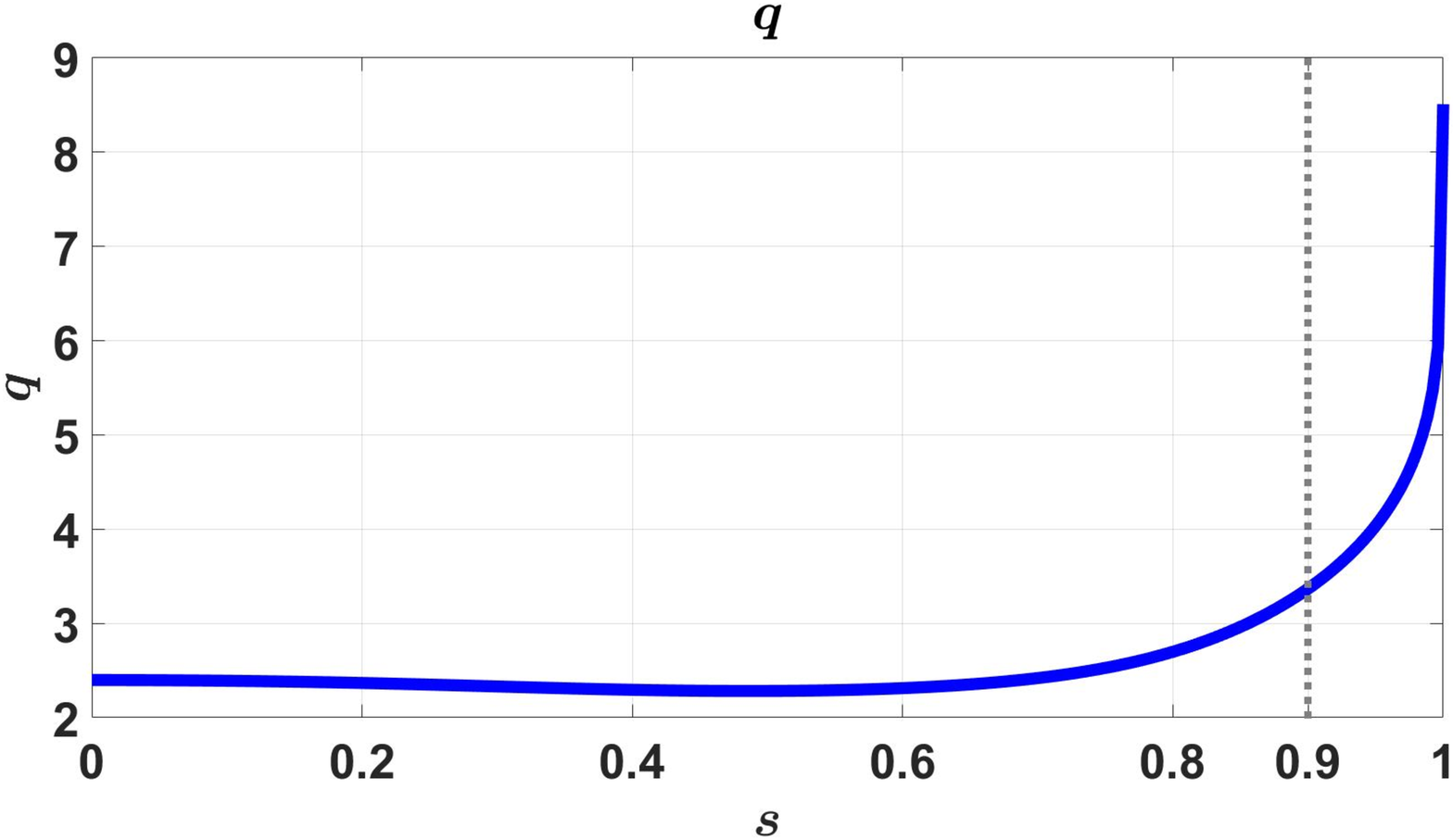}
{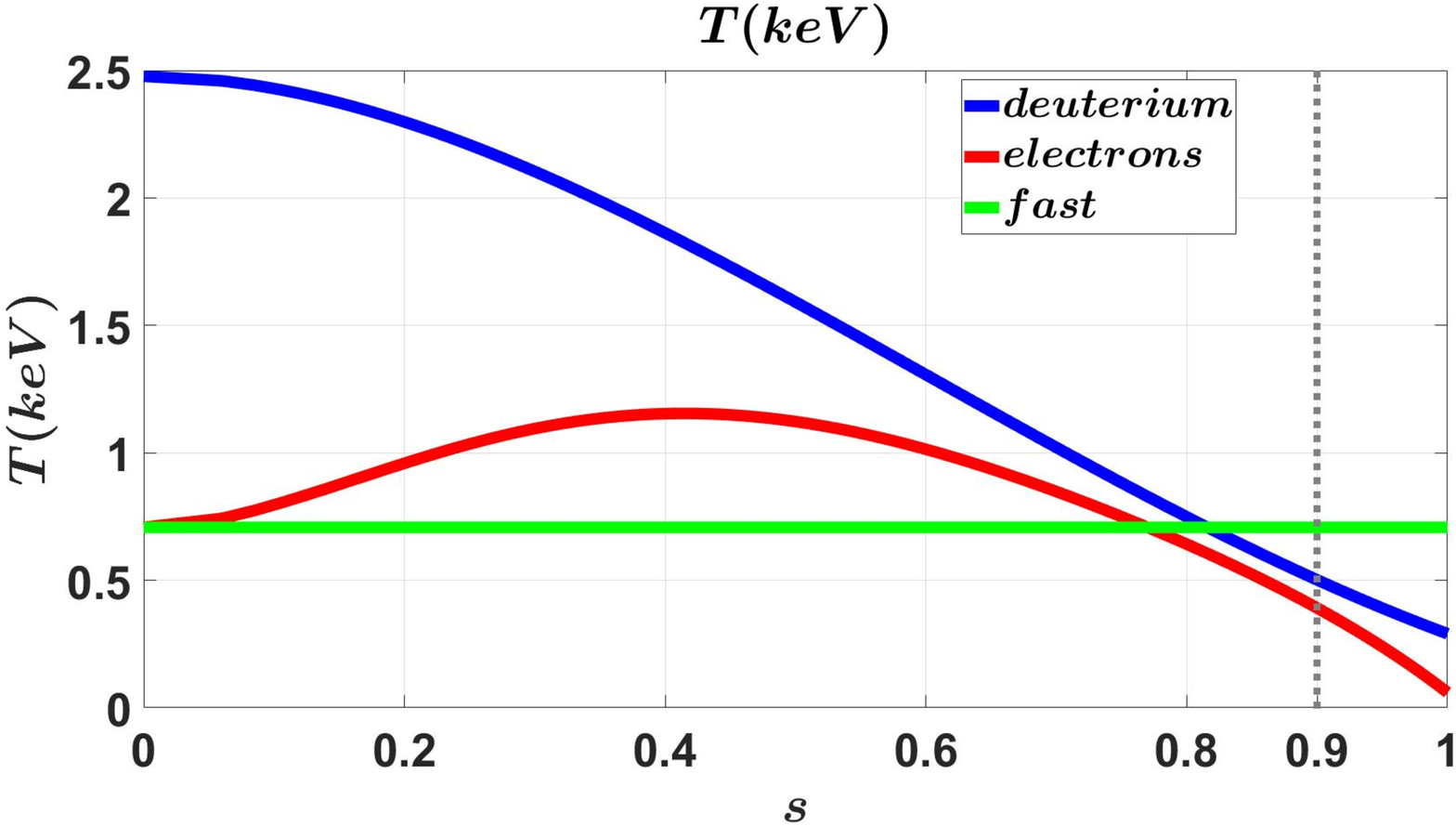}
{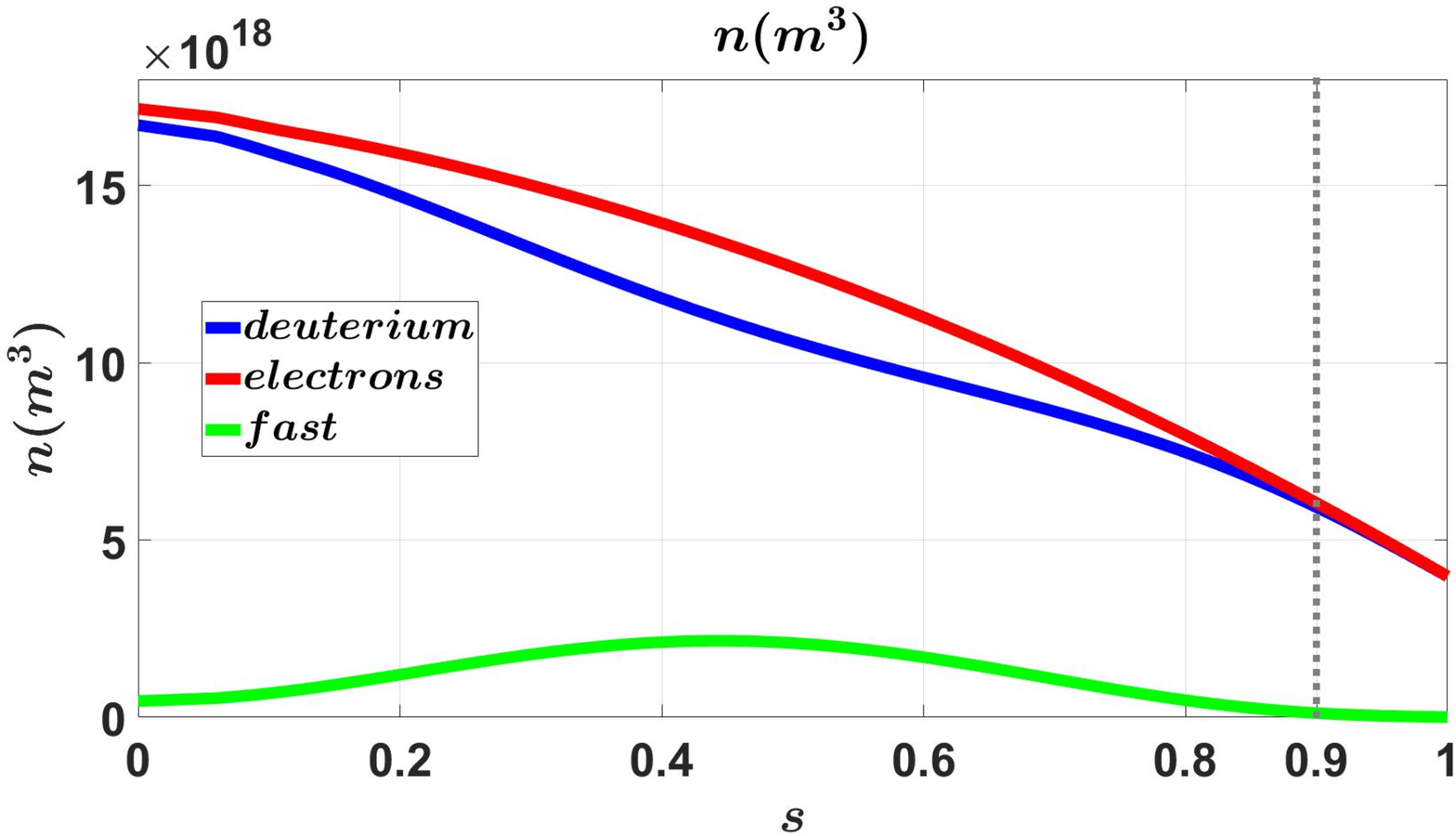}
{Magnetic configuration (upper left plot), radial profile of the safety factor (upper right plot), species temperature (lower left plot) and density (lower right plot) radial profiles for the EGAM simulations in the ASDEX Upgrade shot $\#31213$. The grey vertical dotted lines indicate the right boundary of the simulated radial domain in the EM case with drift-kinetic electrons.}
{fig:nled-equil}  
The real space of the system has been discretized using the following parameters: $n_s = 256$, $n_\chi = 256$, $n_\phi = 32$. In the ES simulation the time grid has a step $dt[\omega_{ci}^{-1}] = 20$ with $N_i = 5\cdot 10^8$ being a number of markers for the thermal ions, and $N_f = 5\cdot 10^8$ for the fast ions.
In the EM case, the time step and number of markers have been changed: $dt[\omega_{ci}^{-1}] = 5$, $N_i = N_f = 10^8$, $N_e = 4\cdot 10^8$. 
Such a high number of markers is needed to provide at least several thousands of numerical markers in every velocity bin, where the mode-plasma resonances are observed (Fig.~\ref{fig:nled-kin-markers}).
In the EM case the radial domain has been reduced to $s = [0.0, 0.9]$ to avoid numerical instabilities due to the abrupt increase of the safety factor at the edge. The density profile, that is depicted in Fig.~\ref{fig:nled-equil}, corresponds to the case with $\beta_e = \langle n_e\rangle T_e/(B_0^2/(2 \mu_0)) = 2.7\cdot 10^{-4}$, where $\langle n_e\rangle$ is the electron density, averaged in a tokamak volume, $\mu_0$ is the magnetic constant, and $T_e$ is measured at the radial position $s = 0.0$.
In both cases, the velocity distribution of the fast particles is described by the expression Eq.~\ref{eq:two-bumps} with $u_{H,f} = 8$ and $T_{H,f} = 1$. 
The ORB5 simulation with such parameters of the fast species results in one of the biggest EGAM growth rate for the given plasma configuration.
\yFigThreeSmall
{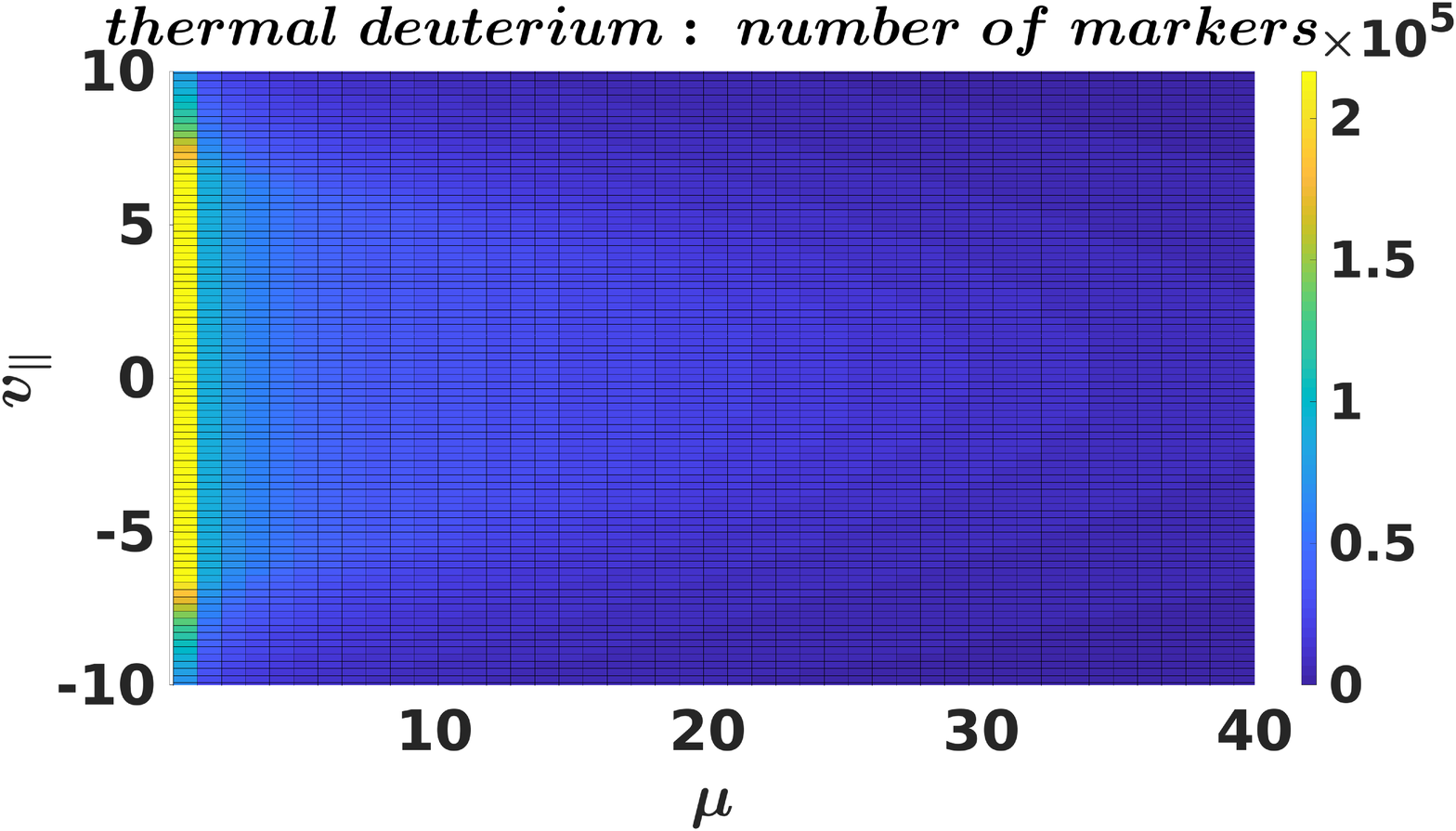}
{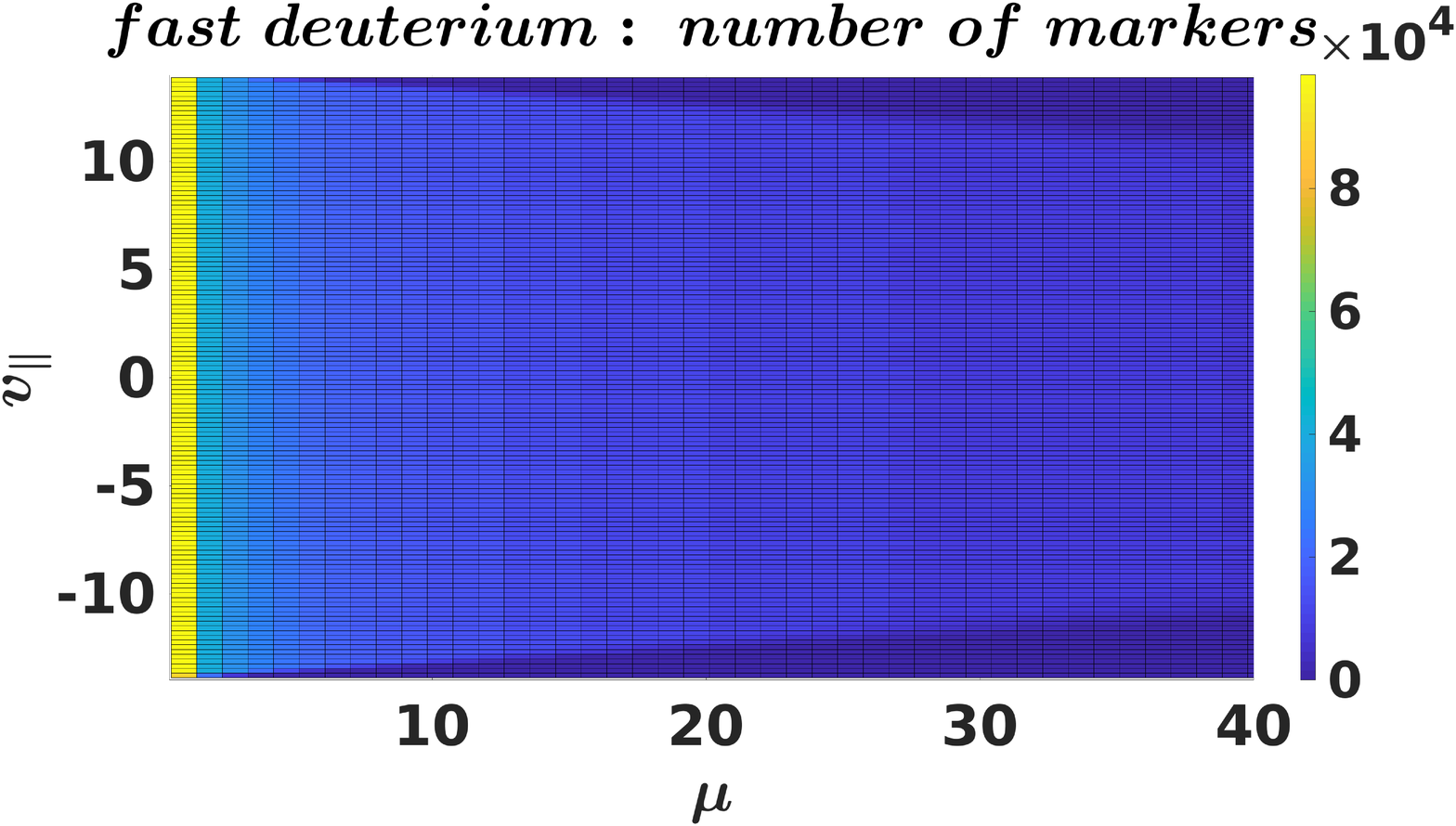}
{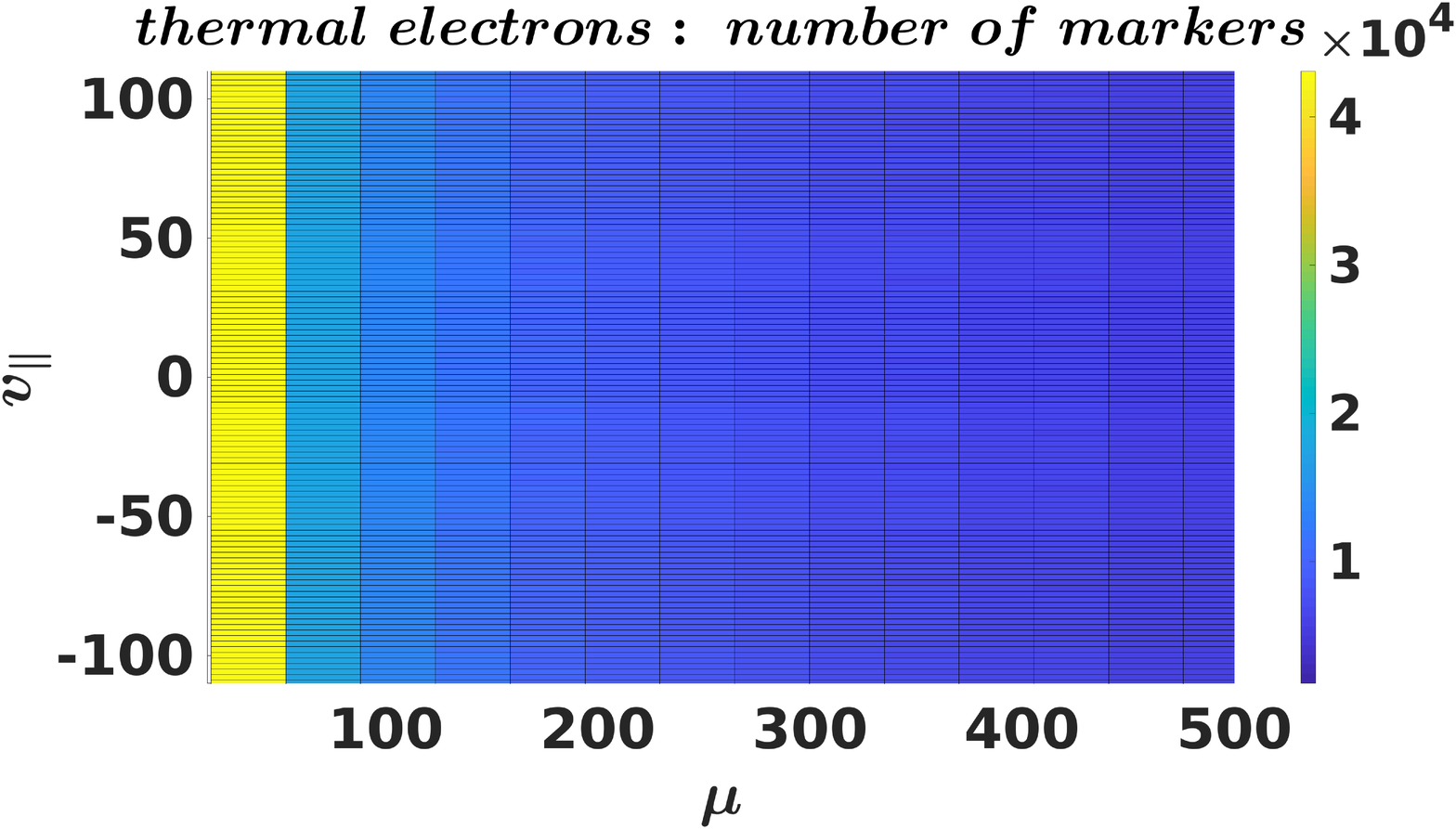}
{Distribution of the numerical markers among velocity bins for different species. The velocity domain is normalized to the sound speed $c_s = \sqrt{T_e(s = 0.0)/m_i}$,  where $m_i$ is a mass of the thermal ion species.}
{fig:nled-kin-markers}

\subsection{Numerical investigation of the wave-particle resonances in the EGAM dynamics}

First of all, one can notice from Fig.~\ref{fig:nled-erbar-st} that the radial structure of the EGAMs slightly changes when the dynamics of the drift-kinetic electrons is switched on. The position of the crest in the EGAM radial structure shifts inwards from around $s = 0.48$ to $s = 0.40$. 
\yFigTwo
{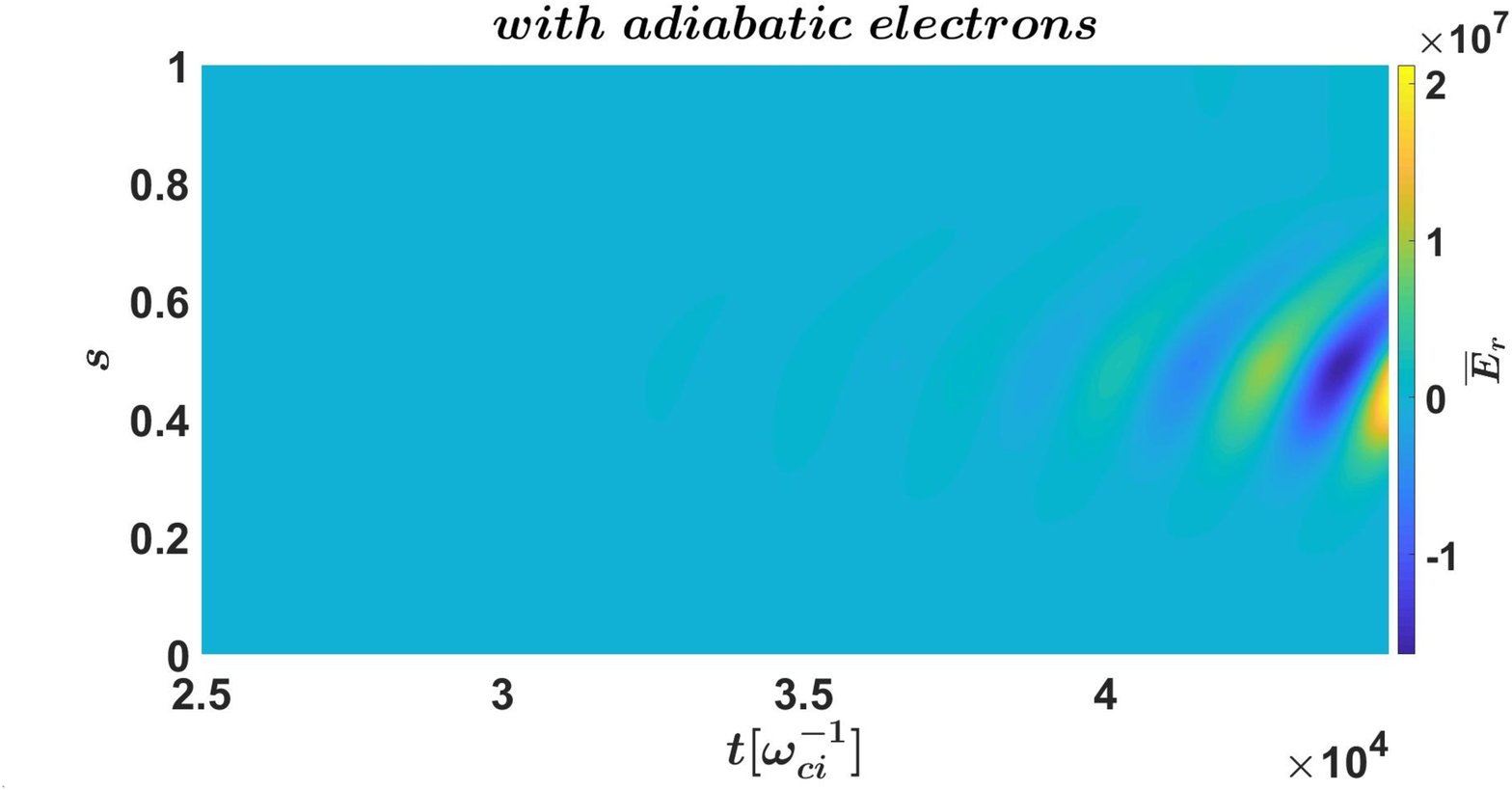}
{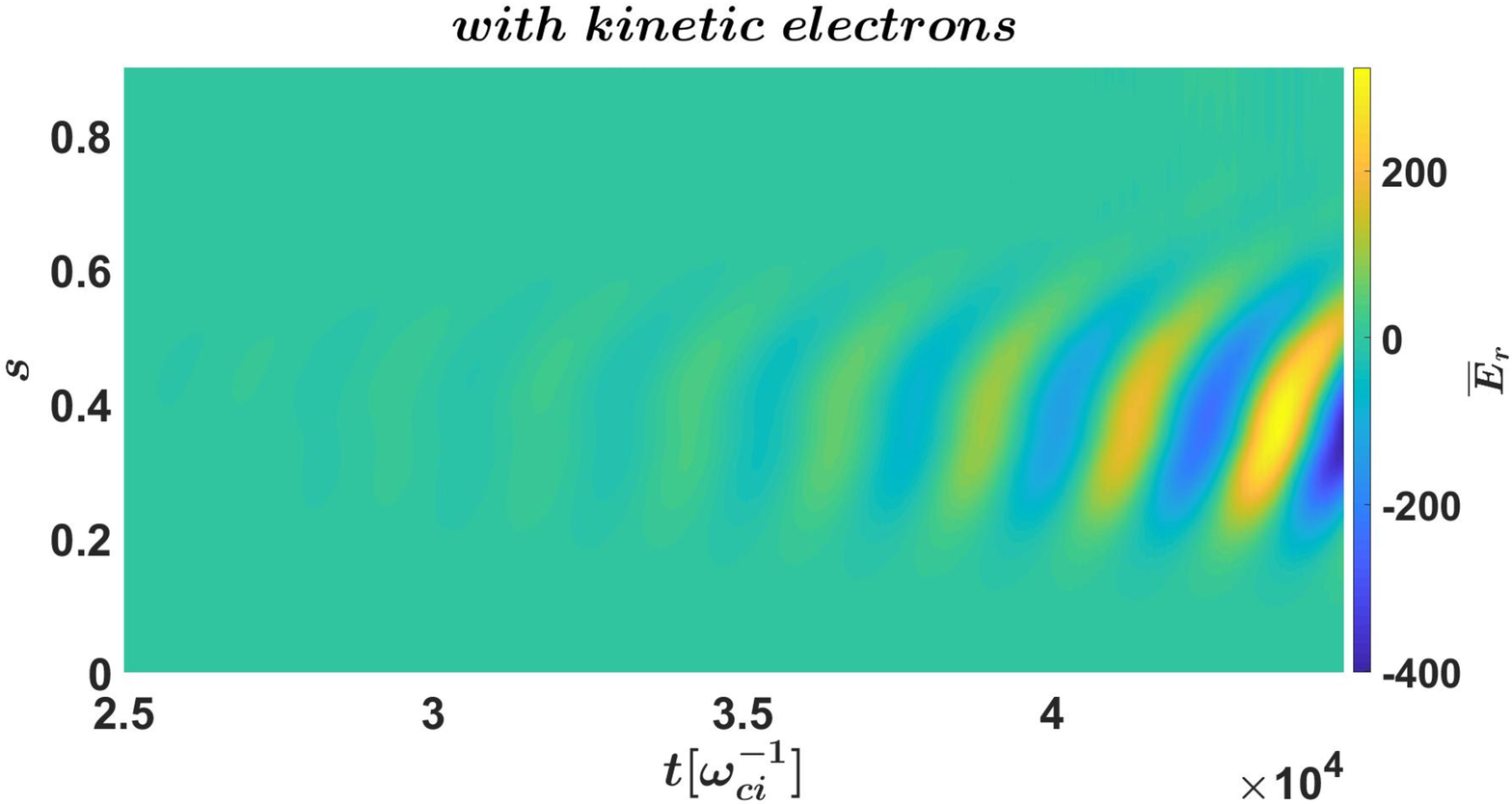}
{Comparison of the radial structure of $\overline{E}_r$ in simulations with adiabatic (left plot) and drift-kinetic (right plot) electrons.}
{fig:nled-erbar-st}
Considering firstly the ES case with AE, we compare the EGAM frequency and growth rate, calculated at radial positions $s = 0.40$ and $s = 0.48$, using the non-linear fitting of $\overline{E}_r$. They appear to be consistent within the error bars of the measurements:
\aeqn
s = 0.40:\ &&\omega\nvR = 9.3\cdot 10^{-1} \pm 2.7\cdot 10^{-2},
				\label{eq:nled-adiab-er040-w}\\
  			 &&\gamma\nvR = 1.6\cdot 10^{-1} \pm 4.7\cdot 10^{-2},
  			 	\label{eq:nled-adiab-er040-g}\\
s = 0.48:\ &&\omega\nvR = 9.20\cdot 10^{-1} \pm 3.1\cdot 10^{-2},
				\label{eq:nled-adiab-er048-w}\\
  			 &&\gamma\nvR = 1.65\cdot 10^{-1} \pm 5.5\cdot 10^{-2}.
  			 	\label{eq:nled-adiab-er048-g}
\eeqn
From here on, only the radial point $s = 0.40$ is considered in the following calculations.
Consistency between the EGAM growth rate, calculated directly from $\overline{E}_r$ and by the MPR diagnostic, using Eq.~\ref{eq:gamma1},
significantly improves in time due to the growth of the EGAM signal in comparison with the zero-frequency zonal flow. 
Skipping initial transient time period, the MPR diagnostic can be applied to measure the EGAM growth rate, that appears to be consistent with Eq.~\ref{eq:nled-adiab-er040-g}:
\aeqn
MPR:\ \gamma\nvR = 1.62\cdot 10^{-1} \pm 1.5\cdot 10^{-3}.
    \label{eq:nled-adiab-mpr-g}
\eeqn
The consistency between both methods is observed in the EM case with KE as well:
\aeqn
\overline{E}_r(s = 0.4):\ &&\omega\nvR = 
				     9.5\cdot 10^{-1} \pm 2.3\cdot 10^{-3},
				     	\label{eq:nled-kin-er04-w}\\
\overline{E}_r(s = 0.4):\ &&\gamma\nvR = 
				     8.3\cdot 10^{-2} \pm 2.9\cdot 10^{-3},
				     	\label{eq:nled-kin-er04-g}\\
MPR:\ &&\gamma\nvR = 8.4\cdot 10^{-2} \pm 9.3\cdot 10^{-3}.
						\label{eq:nled-kin-mpr-g}
\eeqn
From Eq.~\ref{eq:nled-adiab-er040-w} and Eq.~\ref{eq:nled-kin-er04-w} one can see that the change in the EGAM frequency is small in comparison with the change in the growth rate, when dynamics of the drift-kinetic electrons is included. In particular, the EGAM growth rate decreases from Eq.~\ref{eq:nled-adiab-mpr-g} to Eq.~\ref{eq:nled-kin-mpr-g}.
We now want to investigate the role of the drift-kinetic electrons in the EGAM dynamics to understand which wave-particle interactions lead to the decrease of the EGAM total growth rate, by estimation of the contribution of different species. 
In the simulation with adiabatic electrons:
\aeqn
thermal\ deuterium:\ &&\gamma\nvR = -2.99\cdot 10^{-1} \pm 2.3\cdot 10^{-3},
	\label{eq:nled-adiab-mpr-g-i}\\
fast\ deuterium:\ &&\gamma\nvR = 4.62\cdot 10^{-1} \pm 1.3\cdot 10^{-3}.
	\label{eq:nled-adiab-mpr-g-f}
\eeqn
These equations show that the total EGAM growth rate is a balance between the drive on the fast species and damping on the thermal one (one can see also Ref.~\cite{Zarzoso14} for a similar analysis in the case of EGAMs in simplified configurations, with adiabatic electrons). Moreover, the absolute values of the species contributions are much higher than the absolute value of the EGAM total growth rate.

In case with drift-kinetic electrons, the species contributions are the following:
\aeqn
thermal\ deuterium:\  &&\gamma\nvR = -3.8\cdot 10^{-1} \pm 3.2\cdot 10^{-2},
	\label{eq:nled-kin-mpr-g-i}\\
thermal\ electrons:\   &&\gamma\nvR = -3.0\cdot 10^{-2} \pm 9.6\cdot 10^{-4},
	\label{eq:nled-kin-mpr-g-e}\\
fast\ deuterium:\ &&\gamma\nvR =  4.6\cdot 10^{-1} \pm 4.1\cdot 10^{-2}.
	\label{eq:nled-kin-mpr-g-f}
\eeqn
From the above equations one can see that the drive on the fast particles does not change (Eq.~\ref{eq:nled-adiab-mpr-g-f} and \ref{eq:nled-kin-mpr-g-f}). On the other hand, there is a significant increase of the EGAM damping on the thermal deuterium plasma (Eq.~\ref{eq:nled-adiab-mpr-g-i} and \ref{eq:nled-kin-mpr-g-i}). Since this increase is comparable with the electron contribution (Eq.~\ref{eq:nled-kin-mpr-g-e}), one can not claim from these results that the decrease of the EGAM growth rate occurs only directly due to the additional damping on electrons. But apart from the direct damping, inclusion of drift-kinetic electrons changes the position of the EGAM crest 
(Fig.~\ref{fig:nled-erbar-st}) and slightly changes the EGAM frequency. These changes can lead to the increase of the EGAM damping on thermal deuterium. On the other hand, the corresponding errorbars of the ion contributions become higher in the simulation with KE in comparison with AE. Nevertheless, it is clearly shown here that in the experimentally relevant plasma conditions the inclusion of the drift-kinetic electrons significantly decreases the EGAM growth rate of about a factor 2.

We now want to investigate the role of the different resonances in phase space.
In Fig.~\ref{fig:nled-kin-mpr-je-vmu2} one can see the energy transfer signal for the EGAM-electron interaction in the velocity space, averaged on several EGAM periods.
The white cone there indicates an analytical estimation of the boundary between the passing-trapped electrons:
\aeqn
v^{p-tr}_\parallel = \sqrt{2\epsilon \mu},\label{eq:p-tr-bound}
\eeqn
where $\epsilon$ is an inverse aspect ratio.
According to that figure, the EGAMs are damped by the electrons which are localised mainly near this boundary, similar to what happens for 
GAMs~\cite{Zhang10}. 
We can separate two velocity domains $e11$ and $e21$, shown in Fig.~\ref{fig:nled-kin-mpr-je-vmu2}. 
The resonances in the domain $e11$ correspond to the EGAM interaction with the barely trapped electrons, while the domain $e21$ corresponds to the EGAM damping on the barely passing electrons. 
By averaging in the chosen velocity domains, one gets the time evolution of the energy transfer signal (Fig.~\ref{fig:nled-kin-mpr-je-vmu1}), that should be filtered for its proper use in Eq.~\ref{eq:gamma1}.
According to the MPR diagnostic, the contribution of the barely trapped electrons is much more significant than that of the barely passing electrons as one can see from Eq.~\ref{eq:mpt-e11} and \ref{eq:mpt-e21}:
\aeqn
e11:\ &&\gamma\nvR = -1.02\cdot 10^{-2} \pm 2.1\cdot 10^{-4},
\label{eq:mpt-e11}\\ 
e12:\ &&\gamma\nvR = -1.22\cdot 10^{-2} \pm 3.4\cdot 10^{-4},
\label{eq:mpt-e12}\\ 
e13:\ &&\gamma\nvR = -1.50\cdot 10^{-2} \pm 4.8\cdot 10^{-4},
\label{eq:mpt-e13}\\
e14:\ &&\gamma\nvR = -2.1\cdot  10^{-2} \pm 1.1\cdot 10^{-3},
\label{eq:mpt-e14}\\
e21:\ &&\gamma\nvR = -1.19\cdot  10^{-3} \pm 3.9\cdot 10^{-5}.
\label{eq:mpt-e21}
\eeqn
On the other hand, if we consider wider velocity domains, we can notice that there is still a significant contribution of electrons with higher parallel velocity to the EGAM damping. 
One can see it, for instance, comparing the $e13$ and $e14$ velocity domains 
(Eq.~\ref{eq:mpt-e13} and \ref{eq:mpt-e14}). 
The reason might be in the choice of the velocity space variables in ORB5, which has been explained in Sec.~\ref{sec:MPR} in Eq.~\ref{eq:pz-vp}.
\begin{figure}[!ht]
	\ySubFigSL{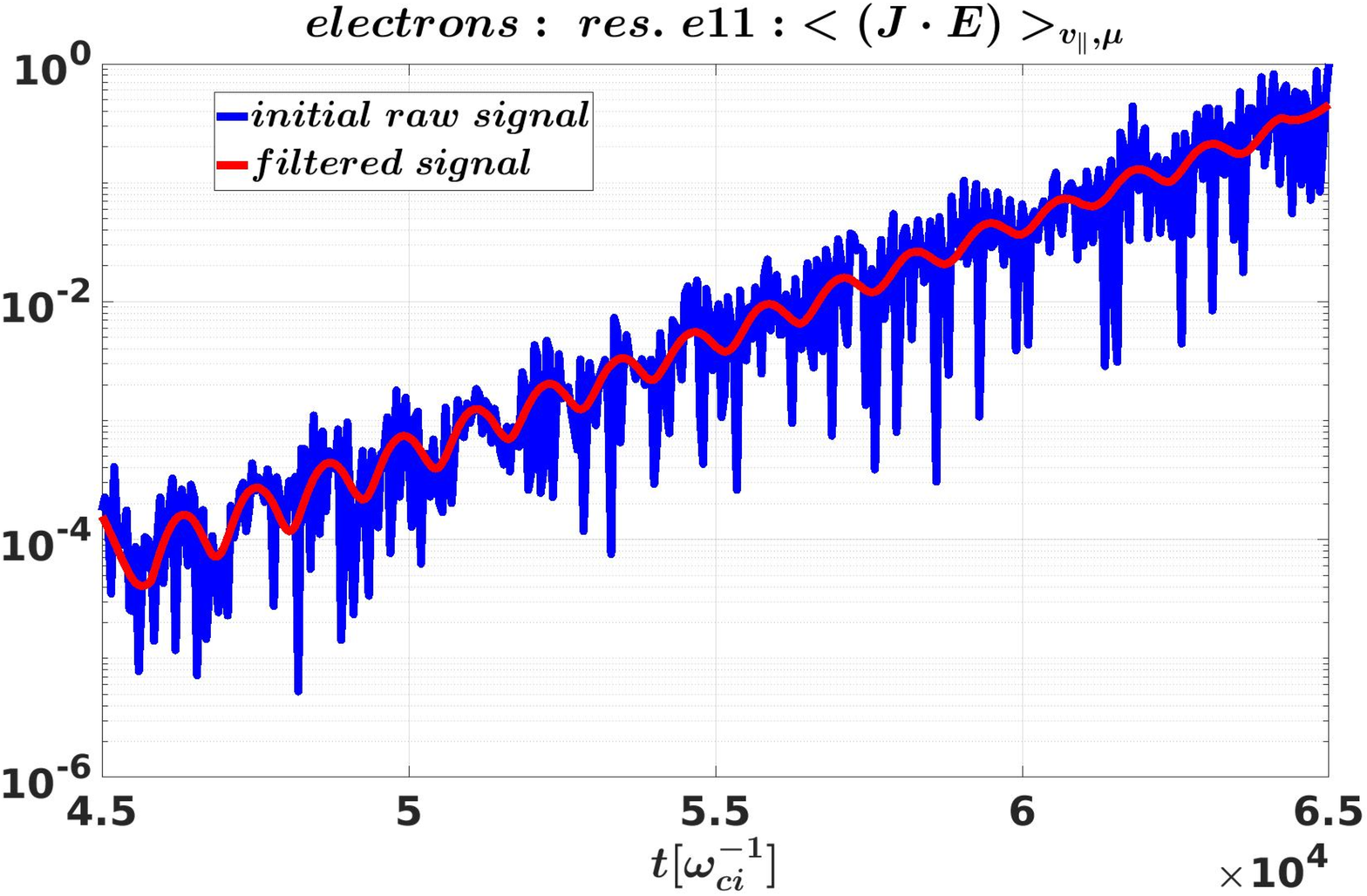}{fig:nled-kin-mpr-je-vmu1}\ySubFigSL{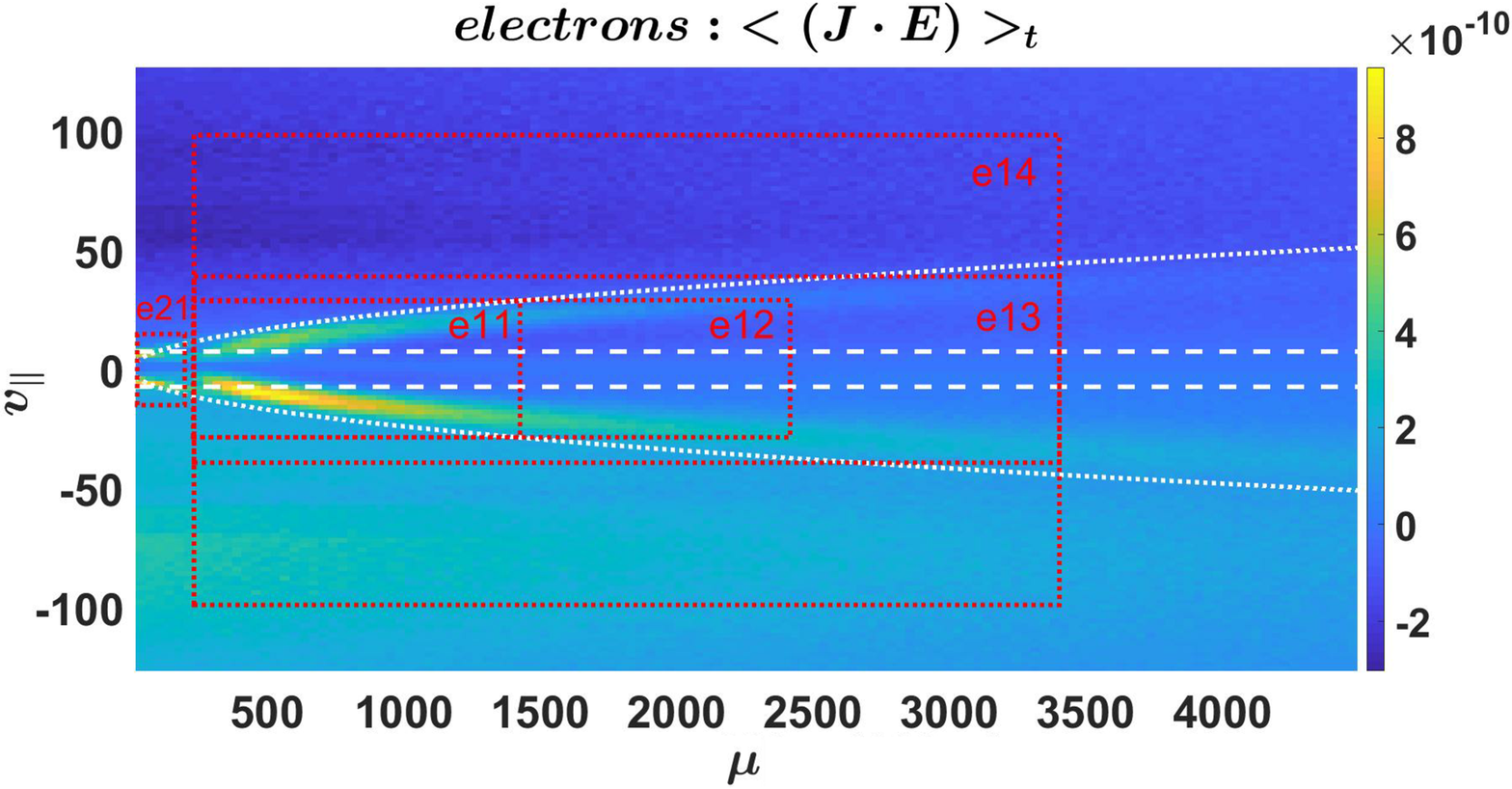}{fig:nled-kin-mpr-je-vmu2}
	\ySubFigSL{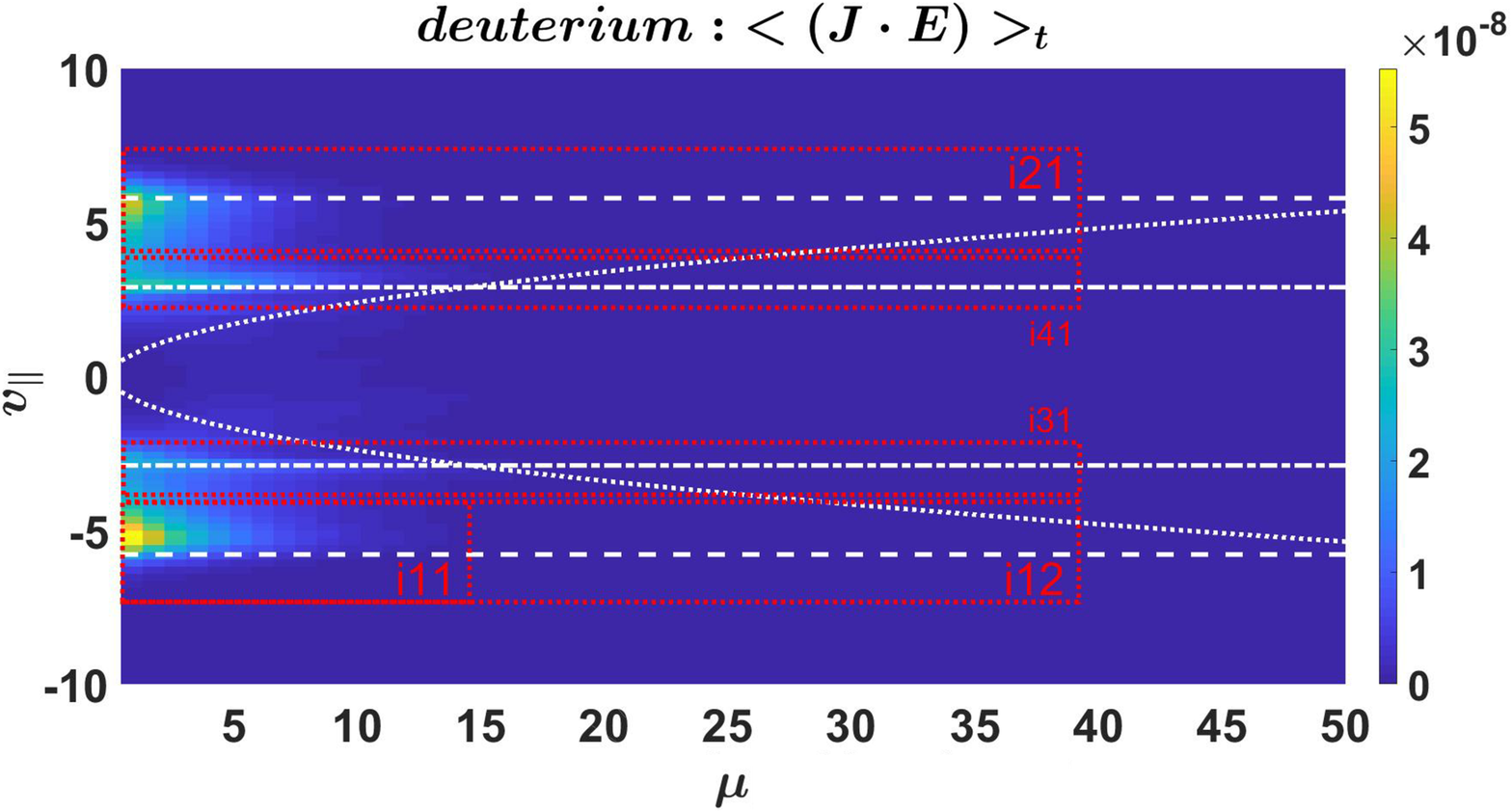}{fig:nled-kin-mpr-je-vmu3}\ySubFigSL{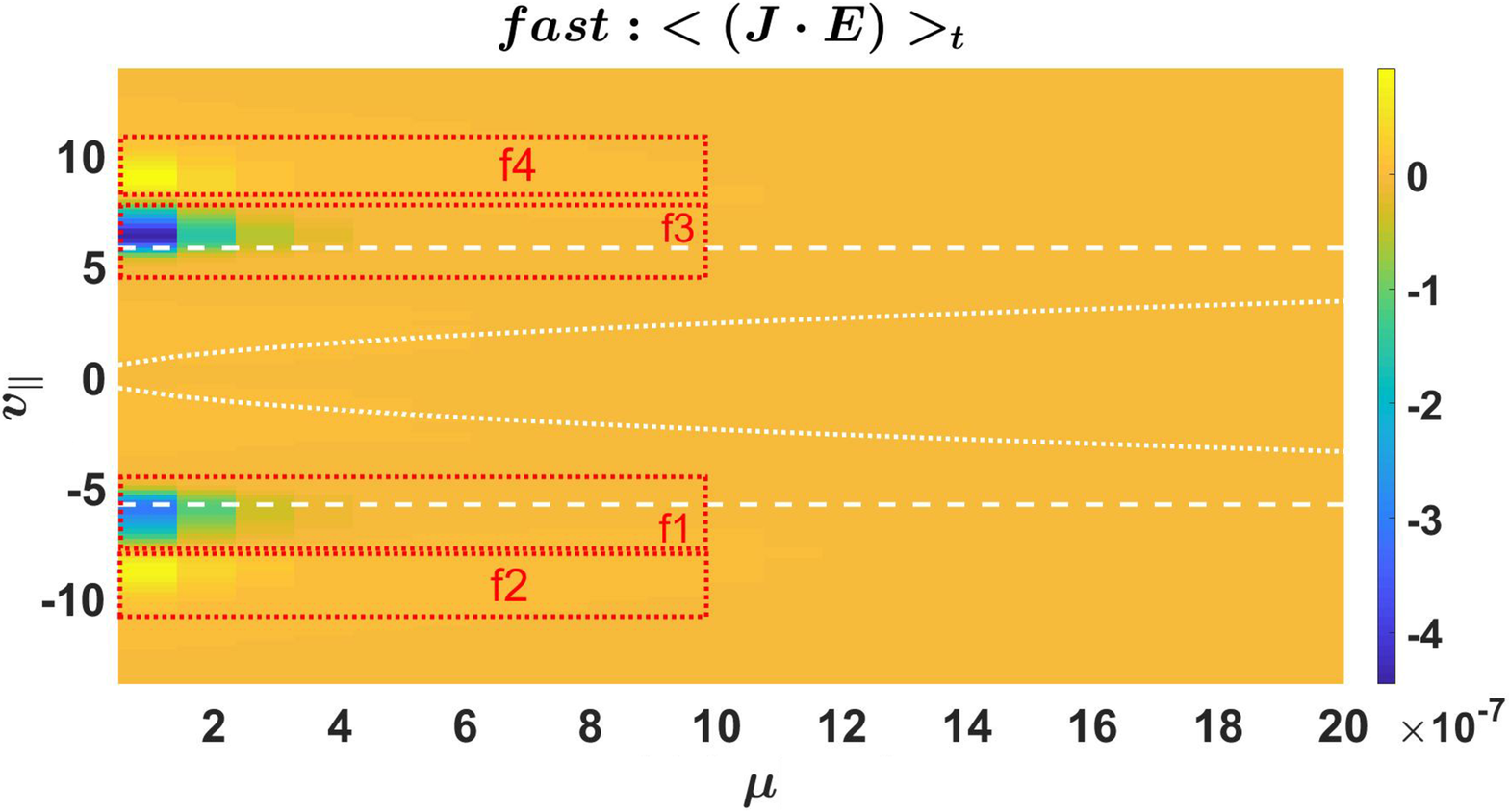}{fig:nled-kin-mpr-je-vmu4}
\caption{Energy transfer signal, averaged in velocity domain $e11$ 
(upper left plot). The blue line corresponds to the initial raw signal, while the red line shows the signal, after low-pass filtering. 
On the upper right plot, the energy transfer signal, averaged on several EGAM periods, is shown with indication of different velocity domains, where the electrons contribution to the EGAM dynamics is investigated.
The energy transfer signals in the velocity space for the thermal deuterium (lower left plot) and fast deuterium (lower right plot) are shown as well. The white dashed lines indicate the analytical estimation of the main EGAM-plasma resonance 
(Eq. \ref{eq:vres}). The dash-dot lines indicate position of the second resonance $v_{\parallel, res}/2$. The white cones indicate analytical estimation of the passing-trapped species boundaries (Eq. \ref{eq:p-tr-bound}). 
\label{fig:nled-kin-mpr-je-vmu}}
\end{figure}
We can estimate as well contribution of different resonances in the thermal deuterium velocity space (Fig.~\ref{fig:nled-kin-mpr-je-vmu3}):
\aeqn
i11:\ &&\gamma\nvR = -1.09\cdot 10^{-1} \pm 4.0\cdot 10^{-3},
\label{eq:i11}\\ 
i12:\ &&\gamma\nvR = -1.10\cdot 10^{-1} \pm 4.3\cdot 10^{-3},
\label{eq:i12}\\
i21:\ &&\gamma\nvR = -8.2\cdot 10^{-2} \pm 2.8\cdot 10^{-3},\\
i31:\ &&\gamma\nvR = -6.9\cdot 10^{-2} \pm 7.8\cdot 10^{-3},\\
i41:\ &&\gamma\nvR = -7.5\cdot 10^{-2} \pm 7.8\cdot 10^{-3},\\
&&i12 + i21 + i31 + i41 \approx -3.4\cdot 10^{-1}.\label{eq:sum-i}
\eeqn
In Eq.~\ref{eq:sum-i} the contributions of all considered resonances are summed up. The result value is close enough to the total contribution of the thermal ions to the EGAM damping (Eq.~\ref{eq:nled-kin-mpr-g-i}). From one point of view, it is an additional option to verify the implemented diagnostic. Moreover, it shows the dominant role of the $m = 1$ resonances (Eq. \ref{eq:vresm}), which are localised in the velocity domains $i12$ and $i21$ 
(Fig. \ref{fig:nled-kin-mpr-je-vmu3}), in the EGAM-thermal deuterium interaction. On the other hand, by comparing Eq.~\ref{eq:i11} and \ref{eq:i12}, one can see that the parallel dynamics has a predominant contribution to the energy exchange between the EGAMs and the thermal deuterium plasma.

Finally, we can consider different velocity domains in the EGAM - energetic deuterium interaction (Fig.~\ref{fig:nled-kin-mpr-je-vmu4}): 
\aeqn
f1:\ &&\gamma\nvR =  3.0\cdot 10^{-1} \pm 1.8\cdot 10^{-2},\label{eq:f1}\\ 
f2:\ &&\gamma\nvR = -9.1\cdot 10^{-2} \pm 3.1\cdot 10^{-3},\label{eq:f2}\\
f3:\ &&\gamma\nvR =  3.7\cdot 10^{-1} \pm 2.2\cdot 10^{-3},\label{eq:f3}\\
f4:\ &&\gamma\nvR = -8.9\cdot 10^{-2} \pm 3.2\cdot 10^{-3},\label{eq:f4}\\
&&f1 + f2 + f3 + f4 = 4.9\cdot 10^{-1}\label{eq:f-sum}
\eeqn
One can see that there is an EGAM damping even on the energetic particles (Eq.~\ref{eq:f2} and \ref{eq:f4}). But it is significantly smaller than the dominant drive (Eq.~\ref{eq:f1} and \ref{eq:f3}). Sum on the resonances (Eq.~\ref{eq:f-sum}) indicates that the EGAMs are driven by the fast species and its absolute value is close enough to the total drive, found in Eq.~\ref{eq:nled-kin-mpr-g-f}.

\section{Comparison with GENE}
\label{sec:gene}
To verify some of the results, obtained in Sec. \ref{sec:nled}, we have performed a comparison with the gyrokinetic GENE code, that has a similar diagnostic.
The Gyrokinetic Electromagnetic Numerical Experiment (GENE)~\cite{GENE} is an Eulerian code, which solves the Vlasov-Maxwell system of coupled equations on the phase-space grid $(\yb{R}, v_\parallel, \mu)$ at each time step. 
Here, $\yb{R}$ denotes the gyrocenter position, $v_\parallel$ the velocity component parallel to the magnetic field and $\mu$ the magnetic moment. 
The gyrokinetic description employs an approach based on the study of a distribution function $f_s (\yb{R}, v_\parallel, \mu)$ for each plasma species $(s)$, which contrarily as it is done in a particle-in-cell code as ORB5, is not discretized with markers. 
The distribution function is split, accordingly to the so-called $\delta$-f approach, into a background component $f_{0,s}$ and in a small fluctuating part $f_{1,s}$, i.e.~$f_s = f_{0,s} + f_{1,s}$. The equilibrium distribution function $f_{0,s}$ is usually modelled with a  Maxwellian distribution. 
However, recently, this assumption has been relaxed and more flexible equilibrium distributions can be considered~\cite{Varenna,PoP,EGAM}. In particular, different analytic choices, e.g. slowing down, bi-Maxwellian and bump-on-tail, as well as numerical distributions as obtained from numerical models are supported. While the equilibrium distributions are considered time independent on the turbulent time scales, their perturbed components evolve in time accordingly to the Vlasov equation, which in the linear and electrostatic limit employed throughout this paper reads as (for more details one can see Ref.~\cite{Tobias,Dannert_1,Dannert_2})
%\begin{dmath}
\begin{align}
\frac{\partial f_{1,s}}{\partial t} 
	+ \frac{\mathcal{C}v_{th,s}}{2\mathcal{J}B_0}\left[v_\parallel^2 
		+ \mu B_0 , h_{1,s}\right]_{zv_\parallel} 
	+ \frac{1}{\mathcal{C}}\frac{\partial \bar{\phi}_1}{\partial y} 
		\partial f_{0,s}\notag\\
	+ \frac{T_0}{q_s}\frac{2v_\parallel^2 +\mu B_0}{B_0}
	\left(\mathcal{K}_x \frac{\partial h_{1,s}}{\partial x} 
	+ \mathcal{K}_y \frac{\partial h_{1,s}}{\partial y} \right) = 0
\label{eq:Vlasov_1}
\end{align}
%\end{dmath}
Here, the function $h_{1,s}$ represents the non-adiabatic part of the perturbed distribution function $f_{1,s}$. 
It is defined as $h_{1,s} = f_{1,s} - q\bar{\phi}_1/(B_0 T_{0,s}) \partial f_{0,s} /\partial \mu$. Eq.~\ref{eq:Vlasov_1} is written in the field aligned coordinate system $(x,y,z)$ with $x$ the radial, $y$ the bi-normal and $z$ the field aligned directions.
Moreover, $\bar{\phi}_1$ denotes the gyro-averaged electrostatic potential, $\mathcal{J}$ the phase-space jacobian, $\mathcal{K}_x \sim - \partial_y B_0 - \partial z B_0$ and $\mathcal{K}_y \sim \partial_x B_0 - \partial_z B_0$, respectively, the radial and bi-normal curvature terms and $\mathcal{C}^2 = \textbf{B}_0 \cdot \textbf{B}_0$. 
Finally, the Poisson brackets are defined as
\aeqn
\left[a ,b\right]_{c,d} = \frac{\partial a}{\partial c} 
	\frac{\partial b}{\partial d} - \frac{\partial a}{\partial d}
	\frac{\partial b}{\partial c}.
\eeqn
Eq.~\ref{eq:Vlasov_1} needs to be solved self-consistently with the Poisson field equation. The full plasma dynamic can be investigated in GENE either in a flux-tube (local assumption)~\cite{GENE} or in a full-global radial domain~\cite{Tobias}. The local approximation allows the radial direction to be Fourier transformed by assuming periodic boundary conditions. 
GENE is able to study the contribution of each plasma species to the overall more unstable mode-dynamic through the study of the time evolution of the potential energy of the system 
$E_w$~\cite{Banon_Navarro_PoP2011,Navarro_PRL2011}. It is defined only in Fourier space (only in the local flux-tube limit) for each wave vector $\textbf{k} = (k_x, k_y)$ as follows
\aeqn
E_w = \left\langle \int d\mu dv_{\parallel}\frac{\pi}{2}B_{0}n_{0}q\bar{\Phi}^*_{1,k}f_{1,k}\right\rangle _{z}.
\label{eq:field}
\eeqn
Here, the bracket represents the field-aligned z-average, namely
\aeqn
\left\langle A\left(z\right)\right\rangle _{z}=\frac{\int \mathcal{J}\left(z\right)A\left(z\right)dz}{\int J\left(z\right)dz}.
\label{eq:ave}
\eeqn
The time derivative of Eq.~\ref{eq:field} determines the energy flow during the whole simulation time domain. In particular it represents the energy effectively transferred from the particles to the field. It reads as
\aeqn
\frac{\partial E_{w}}{\partial t} = \left\langle \int d\mu dv_{\parallel}\frac{\pi}{2}B_{0}n_{0}q\bar{\Phi}_{1,k}^*\partial_{t}f_{1,k}\right\rangle _{z}.
\label{eq:time_field}
\eeqn
From the energy relation of Eq.~\ref{eq:time_field} it is to compute the more unstable linear growth rate $\gamma$ through the time variation of the potential energy, as shown in details in 
Ref~\cite{DiSiena_NF2018,Hatzky_PoP2002,Banon_Navarro_PoP2011,Manas_PoP2015}, by the relation
\aeqn
\gamma = \frac{1}{E_w}\sum_s \frac{\partial E_{w,s}}{\partial t}.
\label{eq:gamma_ene}
\eeqn
Eq.~\ref{eq:gamma_ene} allows us to distinguish between the contribution of each species to the total growth rate, by removing the sum over all the species and studying each term separately. Positive (negative) values of $\partial E_{k,s}/\partial t$ indicate that the plasma species considered is giving (taking) energy to (from) the electrostatic field component with a consequent growth (damping) of the mode. Moreover, by studying $\gamma_s$ in phase-space, i.e. $(v_\parallel, \mu)$ for each plasma species, velocity resonances, which are the main drive term of the EGAMs studied in this paper, can be investigated in details.

The same AUG shot has been simulated in GENE in case with adiabatic electrons (one can see also Ref.~\cite{DiSiena_NF2018}), using the flux-tube version of the code at $s = 0.5$. In Fig.~\ref{fig:gene-res} one can see that both ORB5 and GENE give the same positions of the resonances of the EGAM - fast deuterium plasma interaction. 
According to chosen parameters of the fast deuterium distribution function, peaks of the energetic bumps are located at $|v_\parallel| = 8$.
The opposite signs of the same resonances in two codes is explained by the fact that different signs are used in the MPR diagnostic in ORB5 (Eq.~\ref{eq:gamma1}) and in the corresponding diagnostic in GENE (Eq.~\ref{eq:gamma_ene}). On the other hand, one of the possible explanation of the opposite nature of the resonance asymmetry for positive and negative parallel velocities is the opposite direction of the background magnetic field, used by ORB5 and GENE.
\yFigOne
{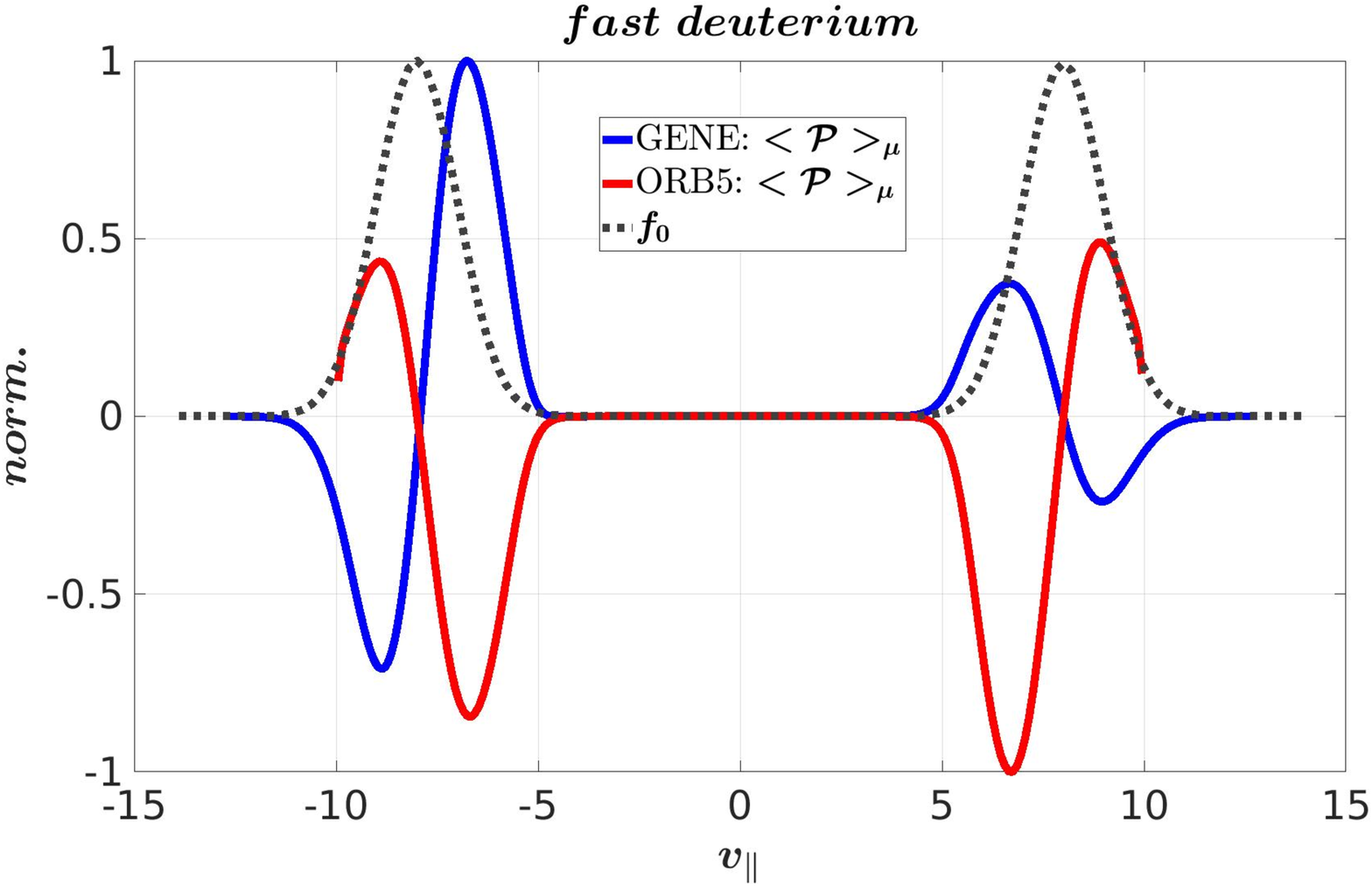}
{Resonance positions of the EGAM - fast deuterium interaction, obtained from GENE and ORB5. The velocity grid is normalised to the sound speed $c_s = \sqrt{T_e(s = 0)/m_i}$. The grey dotted line indicates the positions of the  bumps, which describe the equilibrium distribution of the fast deuterium.}
{fig:gene-res}
As a benchmark, comparison of the EGAM frequency and total growth rate has been done as well. One can see that both codes give the same values of the mode frequency:
\aeqn
GENE:\ &&\omega/2\pi = 42\ (kHz),\label{eq:gene-orb-w}\\
ORB5:\ &&\omega/2\pi = 42.7 \pm 0.1\ (kHz)
\eeqn
On the other hand, there is $18\%$ consistency between the codes for the EGAM total growth rate: 
\aeqn
GENE:\ &&\gamma = 40\cdot 10^3\ (s^{-1}),\label{eq:gene-orb-g}\\
ORB5:\ &&\gamma = (47 \pm 1)\cdot 10^3\ (s^{-1}),
\eeqn
and for the contributions of different plasma species to the mode dynamics:
\aeqn
thermal\ deuterium:\ &&GENE:\ \gamma = -74\cdot 10^{3}\ (s^{-1}),
	\label{eq:gene-orb-g-i}\\
                     &&ORB5:\ \gamma = (-87.6 \pm 0.6)
                     				\cdot 10^{3}\ (s^{-1}),\\
fast\ deuterium:\ &&GENE:\ \gamma = 115\cdot 10^{3}\ (s^{-1}),
	\label{eq:gene-orb-g-f}\\
                  &&ORB5:\ \gamma = (134.8 \pm 0.4)\cdot 10^{3}\ (s^{-1})                    
\eeqn
The difference in the values can be explained mainly by the fact that the simulation in GENE has been performed using the local flux-tube version, while the simulation in ORB5 is a global one.

\section{Conclusions}
\label{sec:conclusions}
In this paper a Mode-Particle-Resonance (MPR) diagnostic has been implemented in the gyrokinetic code ORB5 to investigate mode-plasma interaction processes. 
The technique is based on the projection of energy transfer terms on the velocity space (Eq.~\ref{eq:jdote-disc}) and gives an opportunity to localise velocity domains of maximum energy exchange between an electrostatic mode and different species. 
Moreover, integrating in a chosen velocity domain, a rate of the mode damping or growth can be calculated using Eq.~\ref{eq:gamma1} and contribution of different species to the mode dynamics can be estimated as well.
Using a GAM dispersion relation, which neglects finite-Larmor-radius and finite-orbit-width effects and treats the electrons as 
adiabatic~\cite{Zonca96, Zonca08}, the theoretical principle, which lies behind the MPR diagnostic, has been analytically verified for an ES case. 
It has been shown that the GAM damping rate, derived from the energy exchange principle (Eq.~\ref{eq:gamma1}), is identical to the GAM damping rate, given by the GAM dispersion relation Eq.~\ref{eq:ZoncaDisp}. 
Analytical time evolution of the energy transfer signal, given in Eq.~\ref{eq:je}, has been found to have the same frequency as the numerical one (Fig.~\ref{fig:theory-je}).

In Sec. \ref{sec:nled}, the MPR diagnostic has been applied to the case of AUG shot $\#31213$ (NLED AUG base case) to investigate contributions of different resonances to the EGAM dynamics. It has been shown that inclusion of the drift-kinetic electrons significantly decreases the EGAM growth rate
of about a factor 2 for the selected case (Eq.~\ref{eq:nled-adiab-mpr-g} and Eq.~\ref{eq:nled-kin-mpr-g}). 
It has been shown that the EGAM damping occurs at the first resonance $v_{\parallel, res}$ in case of the interaction with deuterium plasma. On the other hand, in case of the electrons the EGAMs are damped mainly by the barely trapped electrons (Fig.~\ref{fig:nled-kin-mpr-je-vmu}).
The total EGAM growth rate (Eq.~\ref{eq:gene-orb-g}) and contribution of the thermal (Eq.~\ref{eq:gene-orb-g-i}) and energetic deuterium (Eq.
\ref{eq:gene-orb-g-f}) to the mode dynamics has been calculated in the codes ORB5 and GENE in case with adiabatic electrons. The benchmark has shown $18\%$-consistency for the total growth rate and species contributions.

From the point of view of further possible application, the MPR diagnostic can be used, for instance, to study the energy exchange between energetic and thermal species indirectly through the zonal waves, such as EGAMs, that play a role of a mediator in this case. 
Other interesting effects, associated with the EGAM nonlinear evolution, are the EGAM frequency chirping, which consists in a fast modification of the mode frequency, and the saturation mechanisms. 
Since the frequency shift during the chirping is considered to occur as a result of the wave-particle interaction~\cite{Berk97, Berk99, Berk06, Berk10, HWang13, BiancalaniJPP17}, the MPR diagnostic can be used to investigate this phenomenon as well. The saturation mechanisms (wave-particle or wave-wave interactions) are important to investigate in order to build a theoretical model capable of predicting the saturation levels in experimentally relevant conditions, and as a consequence, the EP redistribution in phase space.

The current version of the diagnostic can be applied only to the case of mainly electrostatic modes, such as GAMs and EGAMs. As it has been discussed in 
Sec.~\ref{sec:MPR}, the reason is in the choice of the velocity space variables in ORB5. The MPR diagnostic can be extended to work with EM simulations with arbitrary $\beta$, by performing a proper transition from the variable $p_{z,sp}$ to the velocity variable $v_{\parallel,sp}$.
There are different possible areas of application of the EM-MPR diagnostic. 
A wider range of the modes whose dynamics is mainly controlled by wave-particle resonances, like energetic-particle driven MHD instabilities, can be investigated.
For a turbulent plasma, the collisionless interactions between the EM fields and the plasma particles may lead to a secular transfer of energy from fields to particles, resulting in collisionless damping of the turbulent fluctuations. 
More precisely, a particular challenge in tokamak plasma and plasma physics in general is to identify the physical mechanisms by which the EM field and plasma flow fluctuations are damped and how their energy is converted to plasma heat, or some other energization of particles.
For example, in astrophysical plasmas dissipation of the turbulence energy through the Landau damping of the Alfv\'en waves can take 
place~\cite{Li16}.
It would be interesting to investigate the influence of the plasma $\beta$ on the energy channeling~\cite{Parashar18}, especially on the contributions of different species in the plasma heating by EGAMs and Alfv\'en waves. 
There are also physical phenomena, which are specific to the space plasma, such as particle acceleration by the magnetic energy released during collisions of the magnetic islands in solar and heliospheric environments~\cite{Du18}. It might be interesting to investigate role of such processes in tokamak plasmas as well.

\section*{Acknowledgement}
This work has been carried out within the framework of the EUROfusion Consortium and has received funding from the Euratom research and training program 2014-2018 and 2019-2020 under grant agreement N$^o$ 633053. 
The views and opinions expressed herein do not necessarily reflect those of the European Commission.

Simulations, presented in this work, have been performed on the CINECA Marconi supercomputer within the framework of the OrbZONE and ORBFAST projects.

Stimulating discussions with X. Garbet and V. Grandgirard on the role of kinetic electrons in the GAM dynamics are kindly acknowledged.
One of the authors, I. Novikau, would like to thank F. Vannini for useful conversations. 
One of the authors, A. Biancalani, also wishes to acknowledge stimulating discussions with F. Zonca and Z. Qiu on the GAM/EGAM analytical theory.

%Stimulating discussions with X. Garbet and V. Grandgirard on the role of kinetic electrons in the GAM dynamics are kindly acknowledged. One of the authors, A. Biancalani, also wishes to acknowledge stimulating discussions with S. Briguglio on the study of wave-particle resonances with hamiltonian mapping techniques, which triggered our interest in this work, and with F. Zonca and Z. Qiu on the GAM/EGAM analytical theory.

\bibliography{./MPR-bibl}

\end{document}